\documentclass{aa}
\usepackage{xcolor}
\usepackage{placeins}
\usepackage{stfloats}
\usepackage{graphicx}
\graphicspath{{figures/}}
\usepackage{hyperref}
\usepackage{txfonts}
\usepackage{natbib}
\defcitealias{megeath12}{M12}

\makeatletter
\renewcommand*\aa@pageof{, page \thepage{} of \pageref*{LastPage}}
\makeatother

\newcommand\Tstrut{\rule{0pt}{2.6ex}}         
\newcommand\Bstrut{\rule[-1.2ex]{0pt}{0pt}}   

\begin{document}

\title{Survey of Orion Disks with ALMA (SODA)\thanks{Table 1 is only available in electronic form at the CDS via anonymous ftp to cdsarc.u-strasbg.fr (130.79.128.5) or via \url{http://cdsweb.u-strasbg.fr/cgi-bin/qcat?J/A+A/}}}
\subtitle{I: Cloud-level demographics of 873 protoplanetary disks}

\author{S.E.~van Terwisga\inst{\ref{mpia}}\and A. Hacar\inst{\ref{vienna}}\and E.F. van Dishoeck\inst{\ref{leiden},\ref{garching}}\and R. Oonk\inst{\ref{surf},\ref{leiden},\ref{astron}}\and S. Portegies Zwart\inst{\ref{leiden}}}

\institute{Max-Planck-Institut f{\"u}r Astronomie, K{\"o}nigstuhl 17, 69117 Heidelberg, Germany\label{mpia}~\email{terwisga@mpia.de}\and
		Department of Astrophysics, University of Vienna, T\"{u}rkenschanzstra{\ss}e 17 (Sternwarte) 1180 Wien, Austria\label{vienna}\and
		 Leiden Observatory, Leiden University, PO Box 9513, 2300 RA Leiden, The Netherlands\label{leiden}\and
		 Max-Planck-Institut f{\"u}r Extraterrestrische Physik, Gie{\ss}enbachstraße, D-85741 Garching bei M{\"u}nchen, Germany\label{garching}\and
		 SURF, P.O. Box 94613, NL-2300 RA Leiden, The Netherlands\label{surf}\and
		 Netherlands Institute for Radio Astronomy (ASTRON), Oude Hoogeveensedijk 4, 7991 PD Dwingeloo, The Netherlands\label{astron}
}

\date{\today}

\abstract{Surveys of protoplanetary disks in nearby star-forming regions (SFRs) have provided important information on their demographics. However, due to their sample sizes, these surveys cannot be used to study how disk properties vary with the environment.}
{We conduct a survey of the unresolved millimeter continuum emission of 873 protoplanetary disks identified by {\it Spitzer} in the L1641 and L1647 regions of the Orion A cloud. This is the largest such survey yet, allowing us to identify even weak trends in the median disk mass as a function of position in the cloud and cluster membership. The sample detection rates and median masses are also compared to those of nearby ($< 300$\,pc) SFRs.}
{The sample was observed with the Atacama Large Millimeter/submillimeter Array (ALMA) at 225\,GHz, with a median rms of $0.08$\,mJy\,beam$^{-1}$, or $1.5\,\textrm{M}_{\oplus}$. The data were reduced and imaged using an innovative parallel data processing approach.}
{We detected $58\%$ (502/873) of the observed disks. This includes 20 disks with dust masses $> 100\,\textrm{M}_{\oplus}$, and two objects associated with extended dust emission. By fitting a log-normal distribution to the data, we infer a median disk dust mass in the full sample of $2.2^{+0.2}_{-0.2}\,M_{\oplus}$.  In L1641 and L1647, median dust masses are $2.1^{+0.2}_{-0.2}\,M_{\oplus}$ and $2.6^{+0.4}_{-0.5}\,M_{\oplus}$, respectively.}
{The disk mass distribution of the full sample is similar to that of nearby low-mass SFRs at similar ages of $1-3$\,Myr. We find only weak trends in disk (dust) masses with galactic longitude and between the Young Stellar Object (YSO) clusters identified in the sample, with median masses varying by $\lesssim 50\%$.
Differences in age may explain the median disk mass variations in our subsamples. Apart from this, disk masses are essentially constant at scales of $\sim 100$\,pc. This also suggests that the majority of disks, even in different SFRs, are formed with similar initial masses and evolve at similar rates, assuming no external irradiation, with disk mass loss rates of $\sim 10^{-8}\,\textrm{M}_{\odot}\,\textrm{yr}^{-1}$.} 


\maketitle

\section{Introduction}
The evolution of the dust in protoplanetary disks is an essential part of our understanding of wider questions of planet formation, and for the earliest development of planetary systems. The Atacama Large Millimeter/submillimeter Array (ALMA) has revolutionized the study of this field with its high sensitivity to the emission from cold, millimeter-sized dust grains in the midplanes of these disks~\citep{andrews20}. One way it has done this is by revealing highly structured emission in individual disks at unprecedented resolutions, which show rings, gaps, and spiral arms~\citep[e.g.,][]{hltau,andrews18}. At the same time, the large number of 12m-telescopes in the array has made it possible to perform sensitive surveys typically targeting around a hundred disk-bearing stars at once, with sensitivities down to an Earth mass or less of dust. Unbiased and (nearly) complete surveys have now been performed in many of the nearby low-mass star-forming regions (SFRs)~\citep[e.g.,][]{ansdell16, pascucci16, cieza19,cazzoletti19}, as well as in several of the Orion clouds~\citep[e.g.,][]{mann14,vanterwisga19b,ansdell20,vanterwisga20}. Each of these is usually treated as a snapshot of one set of conditions, and one age bin. Thus, by comparing the continuum emission from disks in different regions, the evolution of dust can be constrained empirically. The observed properties of protoplanetary disks in a given star-forming region seem to be primarily explained by that region's age, and by the presence of strong UV radiation fields from nearby young O-type stars.

In environments where no such young and massive stars are present and where the evolution of dust in disks is thought to be dominated by internal processes, one key result has been that if disks are optically thin at millimeter wavelengths, after 1--3\,Myr, the average disk does not have more than a few Earth masses of dust left for planet formation. This behavior was seen in surveys of disks at submillimeter wavelengths in the Taurus~\citep{andrews13}, Lupus~\citep{ansdell16,ansdell18}, and Chamaeleon I~\citep{pascucci16} SFRs, as well as (more recently) the Corona Australis~\citep{cazzoletti19} and $\rho$ Ophiuchus~\citep{cieza19, williams19} regions. At later times (6--10\,Myr), as exemplified by the survey of Upper Sco disks~\citep{barenfeld16}, the evolution of dust has proceeded even further and led to even lower masses of dust in the median disk than in the previous regions. A highly similar distribution of disk masses was observed in the $\sim 5$\,Myr-old $\lambda$ Ori region~\citep{ansdell20}.

On the other hand, the presence of O-type stars can have a dramatic effect on protoplanetary disks. In the young (1--3\,Myr) Trapezium region, clear evidence of an increasing disk mass with distance from the cluster center has been reported~\citep[e.g.,][]{mann14, eisner18}. Beyond $\sim 2$\,pc from the Trapezium, however, disk masses are similar to those in Lupus and Taurus~\citep{vanterwisga19b}. A strong gradient in disk masses was also seen in $\sigma$ Ori~\citep{ansdell17}. This is taken as evidence that protoplanetary disk masses are strongly affected, not only by their age, but by the strong far-ultraviolet radiation of nearby O-stars. In the younger NGC 2024, which contains several massive O-type stars, there is likewise clear evidence that external photoevaporation of protoplanetary disks is important and occurs even as early as 0.5 -- 1\,Myr~\citep{vanterwisga20,haworth21}.

Some key questions, however, remain open, despite these surveys' important results. In particular, it is difficult to say if disks are formed with some universal initial mass  distribution, which then evolves similarly in the absence of external photoevaporation, or if different star-forming regions have different initial disk properties or mass loss rates even if no significant external photoevaporation is expected. The strong similarities between the disk mass distributions of approximately equal-age Lupus, Taurus and OMC-2 SFRs, and between the older Upper Sco and $\sigma$ Ori populations, supports this. However, $\rho$ Oph and CrA are also supposed to be young (comparable in age to Taurus, or, in the case of $\rho$ Oph, even younger) and have significantly lower median disk masses and different disk mass distributions: in CrA and $\rho$ Oph, $\textrm{M}_{\textrm{dust, median}} = 0.3^{+0.1}_{-0.2}\,\textrm{M}_{\oplus}$ and $0.8^{+0.1}_{-0.2}\,\textrm{M}_{\oplus}$, respectively, compared to $3.3^{+0.5}_{-0.6}\,\textrm{M}_{\oplus}$ in Taurus. At present, it is not clear to what extent this result may be caused by the underlying stellar age distributions in these regions~\citep{cazzoletti19,galli20}.

This problem is exacerbated by the relatively small number of regions that were studied: even considering the large uncertainties in the ages, the best-populated bin is the 1--3\,Myr old range, which contains a handful of comparable SFRs where no O-type stars are present. An additional complication here is that each region includes a quite wide range of physical conditions, such as stellar densities, but not enough objects to study their impact. To give an example, the Lupus SFR's Class II disk population ($N=98$ Class II disks) is distributed over four main star-forming clouds, with most of the disk-bearing stars located in the Lupus III cloud. However, the local stellar density in Lupus III and, say, Lupus II is quite different: Lupus III is sharply concentrated, while the disk-bearing stars in Lupus II are spread out throughout the cloud. However, the sample sizes are too small to be treated independently in each separate cloud. The physical limits on complete sample sizes in nearby star-forming regions are thus a key problem in our detailed understanding of the evolution of protoplanetary disks.

Within Orion, the question of the initial disk mass distribution is also important. After all, it provides the fundamental constraints behind attempts to solve the proplyd lifetime problem~\citep{scally01}, and the star-formation history of the Orion Nebula Cluster (ONC), an important open problem~\citep[e.g.,][]{winter19}. In the OMC-2 region, disk masses suggest comparable properties to Lupus and Taurus at similar ages.

The large number of Class II Young Stellar Objects (YSOs) in Orion A (at least 2400 were identified with the {\it Spitzer} Space Telescope~\citealt{megeath12}; hereafter \citetalias{megeath12}) and the sheer size of the cloud, at about 90\,pc in length~\citep{grossschedl18}, means that this region is perfect to study the properties of large samples of disks. In particular, the part of the cloud consisting of L1641 and L1647 is a key region for studying the properties of disks at $\sim 1$\,Myr ages in giant molecular clouds, away from young O-type stars.

Another reason that Orion A so valuable for observations of protoplanetary disks is that there is a wealth of complementary information to facilitate our analysis. Apart from the previously-mentioned {\it Spitzer}-based survey of disks and embedded stars, observations with {\it XMM-Newton} at X-ray wavelengths~\citep[e.g.,][]{pillitteri13} and with {\it Herschel}~\citep{furlan16} mean that both younger and older YSO populations have been mapped. The embedded disks (of Class 0 and Class I-YSOs) have also been characterized by ALMA at millimeter wavelengths by~\citet{tobin20}. The availability of \textit{Gaia} data has been essential for understanding the spatial and velocity structure of the cloud's stellar populations, their ages, and the structure of the Orion A cloud itself in greater detail~\citep{zari19,rezaei20}. These observations are complemented by accurate very-long-baseline interferometric and spectroscopic surveys like APOGEE-2~\citep{kounkel17,kounkel18}. Given the known correlation between disk and stellar masses~\citep{pascucci16}, it is also important that the stellar mass distribution in Orion A has been well-studied, and found to be similar to that of nearby SFRs in this part of the cloud~\citep{luhman18}.

Recently,~\citet{grant21} surveyed protoplanetary disks in L1641 with ALMA for the first time. Their sample consists of 104 Class II candidates from~\citetalias{megeath12} that were also detected by {\it Herschel} at 70\,$\mu$m. The authors find a median disk mass of $11.1 _{-4.6}^{+32.9}$\,M$_{\oplus}$ for their sample, assuming optically thin emission and the same dust opacity of $\kappa_{\nu = 230\,\textrm{GHz}} = 2.3$\,cm$^{2}$\,g$^{-1}$ used in this paper. As the authors suggest, it is likely that the 70\,$\mu$m-detection requirement introduces a bias toward more massive disks. Taking this into account, it is plausible that Class II disk masses are lower by some amount, with a median disk mass possibly as low as $0.02^{+0.01}_{-0.01}\,\textrm{M}_{\oplus}$, if all unobserved disks in the sample are are treated as nondetections with ALMA and assuming a log-normal disk mass distribution. To resolve this uncertainty, an unbiased survey is needed.

In this article, we present observations from an unbiased survey with ALMA of 873 Class II protoplanetary disks in the L1641 and L1647 clouds, the Survey of Orion Disks with ALMA, or SODA. In Section~\ref{sec:surveydesign}, the survey design is described. Section~\ref{sec:observations} details the observation strategy and data reduction process. In Section~\ref{sec:results}, the first results from this survey are shown, including the detection rates, the unbiased catalog, and the first complete catalog of the most massive disks in L1641 and L1647. We discuss environmental variations in disk masses in Sections~\ref{sec:alongcloud},~\ref{sec:clusters} and~\ref{sec:otherregions}.

\section{Survey design}
\label{sec:surveydesign}

The aims of this project require the selection of a large, unbiased sample, with uniform selection criteria. As most other similar studies, this survey relies on a {\it Spitzer} survey of YSO colors, which is sensitive to the (optically thick) emission from warm dust in the inner disk, conducted by~\citet{megeath12}. Thus, it should be a good indicator of the presence of a circumstellar disk, without being strongly biased to its overall mass. The catalog also includes transition disks, which are characterized by faint or absent emission from the innermost disk parts of the dust disk, but these are not treated separately here. For consistency, we use the same source identifications as in the original paper, abbreviating as [MGM2012] and the number of the object in their catalog.

\begin{figure*}
\centering
   \includegraphics[width=\textwidth]{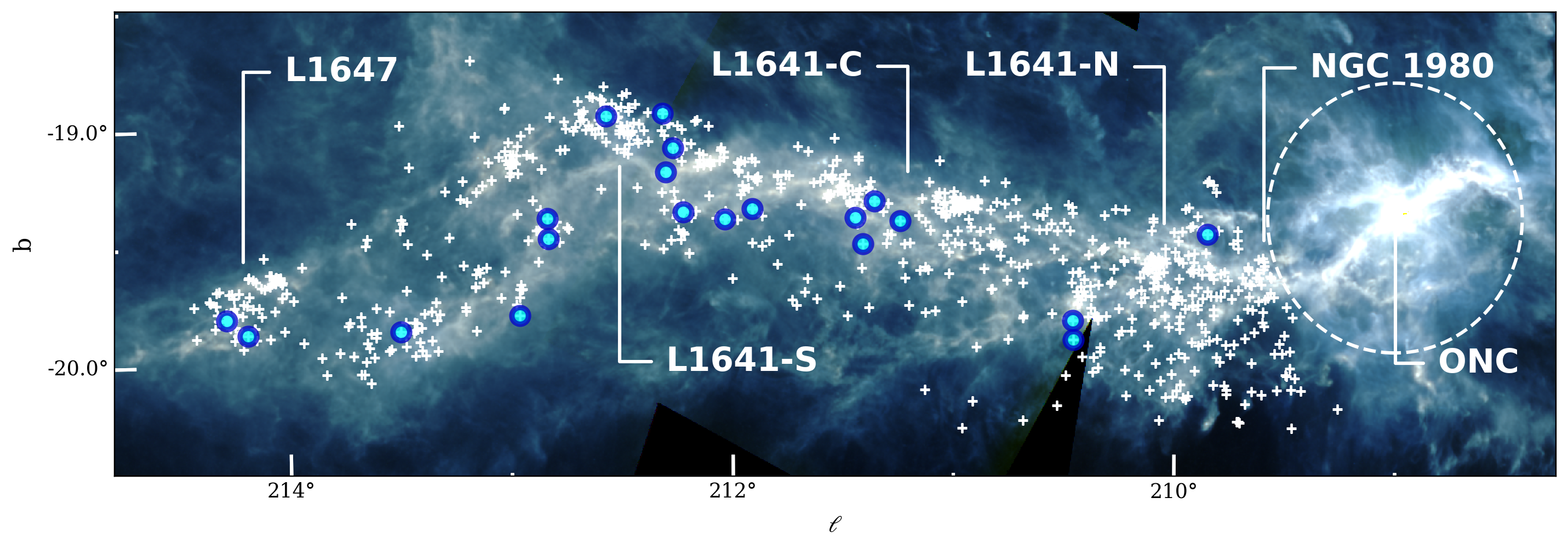}
     \caption{Sample positions on the sky. Disks are marked with white plus signs; disks with masses $\textrm{M}_{\textrm{dust}} > 100 \textrm{M}_{\oplus}$ are shown as blue circles. The major regions within Orion A (L1647, the three subdivisions of L1641, and the NGC 1980 and ONC clusters) are marked separately. {\it Herschel} SPIRE observations at $250, 300,$ and $500\,\mu$m \citep{soler19} form the background. The dashed white circle marks a 4\,pc radius around $\theta^1$ Ori C.}
     \label{fig:pointingmap}
\end{figure*}

	\subsection{Sample selection}
	We selected all Class II YSOs from~\citet{megeath12} located in Orion A and below $-6^{\circ}$ degrees in declination, or above $\ell \sim 209.5^{\circ}$ degrees in Galactic longitude. As Figure~\ref{fig:pointingmap} shows, this simple selection criterion means that we have a large number ($N=873$) of protoplanetary disks in our sample, focused on the L1641 star-forming region. It also ensures we avoid the immediate environment of the ONC, with all of our stars located at least 4\,pc away from the bright stars of the Trapezium. This ensures that their bright UV radiation is not actively influencing the evolution of the disks we are interested in. This sample should thus be relatively similar to other nearby star-forming regions, except for its large number of sources. YSOs in the Orion A cloud are young (but with local variations), with typical mean ages of $\sim 1 - 3$\,Myr~\citep[e.g.,][]{dario16}

	The infrared color criteria used to identify the disk sample can misclassify sources based on their inclination. In particular, edge-on objects might appear to be deeply embedded protostars (or be absent altogether), and face-on embedded sources may appear like more evolved disk-bearing objects~\citep{whitney03,crapsi08}. To correct for this effect, we refined the classifications from~\citet{megeath12} to include the observations from the {\it Herschel} Orion Protostars Survey (HOPS)~\citep{furlan16} at longer wavelengths, targeting embedded sources. From the original sample of 882 disk sources, nine were reclassified to be younger (Class I or Flat-spectrum). We mark these sources in Table~\ref{tab:embedded}. We thus have an 873-source Class II sample.
	
	We expect fewer re-classifications in the other direction. Two of our presumptive Class II objects, [MGM2012] 512 and [MGM2012] 950, have features indicating they may be younger and more embedded sources than typically associated with a Class II Spectral Energy Distribution (SED). Since the evolution of YSOs is a smooth process, even if the transition from Class I to Class II proceeds rapidly when traced with submillimeter luminosity~\citep{tobin20,tychoniec20}, some such ambiguous objects are expected to be present in a large sample like the one presented here, but not at elevated rates compared to other {\it Spitzer}-based disk surveys, and with a limited statistical impact on our conclusions.
	
	In total, after cross-referencing with the HOPS survey, we have measured fluxes (extracted at the phase center of each pointing, assuming unresolved emission) for 873 Class II disks, plus a total of 18 flat-spectrum, 18 Class I, and 4 Class 0 YSOs within the field of view of our observations. The fluxes of the sources that are not considered to be Class II are listed in Table~\ref{tab:embedded}. The larger number of sources than pointings (882) is caused by size of the primary beam of our observations. While no disks were observed outside the primary beam, we could target some additional embedded objects. This also means that the younger YSOs have more uncertain photometry, on the whole, due to the decrease in signal to noise away from the primary beam. In the following, we focus on the results for the Class II objects only.

	\subsection{Catalog completeness}
	The completeness of the {\it Spitzer}-survey has been thoroughly characterized by~\citet{megeath16}. Due to the presence of strong background emission in some parts of Orion, such as the ONC, {\it Spitzer} is less sensitive to fainter stars there (which have lower masses). This can, in turn, introduce a bias in our ALMA survey: lower-mass stars bear fainter and lower-mass disks, although with significant scatter~\citep[e.g.,][]{ansdell16, barenfeld16, pascucci16}. However,~\citet{megeath16} conclude that the completeness fraction is high in regions of faint nebulosity, and where stellar surface densities are low. This applies in particular to L1641 and L1647, where the highest completeness is reached, of $> 80\%$. This is also the region targeted by this survey, and indicates that completeness is not a concern for the area covered by SODA. This is consistent with the results from~\citet{grossschedl19}, who added a relatively small number of disk-bearing sources throughout the Orion A cloud, but did not find large differences from the {\it Spitzer}-based catalog. Additionally, while in the brightest parts of the ONC the 24\,$\mu$m band saturates, this is not a concern for the area covered by this survey \citep{megeath12}.
	
	Finally, we must consider the possiblity that some of the objects in the sample are not YSOs at all. This is the most difficult category of error to quantify. The primary candidates for such false-positive objects in the sample are background galaxies\citep{gutermuth08}, IR nebulae associated with YSOs, but not coinciding with the protostar itself within our search radius, and in-telescope effects like diffraction spikes. Since the color criteria used in this catalog are similar to those used to select disk-bearing stars in nearby SFRs, these contamination rates should also be similar across the literature. We do not expect to detect any of the galaxies possibly contaminating our ALMA observations~\citep[e.g.,][]{casey14}. Likewise, an IR nebula associated with a protostar can lead to an offset from the source position that is large enough to lead to a nondetection. The inclusion of these types of objects will therefore always lead us to underestimate of the median disk luminosity. The galaxy contamination fraction of the catalog is expected to be approximately $6.1 \pm 3.1$\,deg$^{-2}$, or $24 \pm 12$ galaxies in total by~\citet{megeath12} for the area surveyed here and the color cuts employed. In the analysis of the YSO population in Orion A by~\citet{grossschedl19}, the authors infer a similarly low contamination rate for these contaminants in their Class II sample spanning the same area on the sky, of $\sim 2.3\%$, based on the visual identification of contaminants in near-infrared data from the Visible and Infrared Survey Telescope for Astronomy (VISTA)~\citep[see also][]{meingast16}. In any case, contamination should not be a dominant source of nondetections in this survey.

\section{Observations and data reduction}
\label{sec:observations}
\begin{figure}
\centering
   \includegraphics[width=\linewidth]{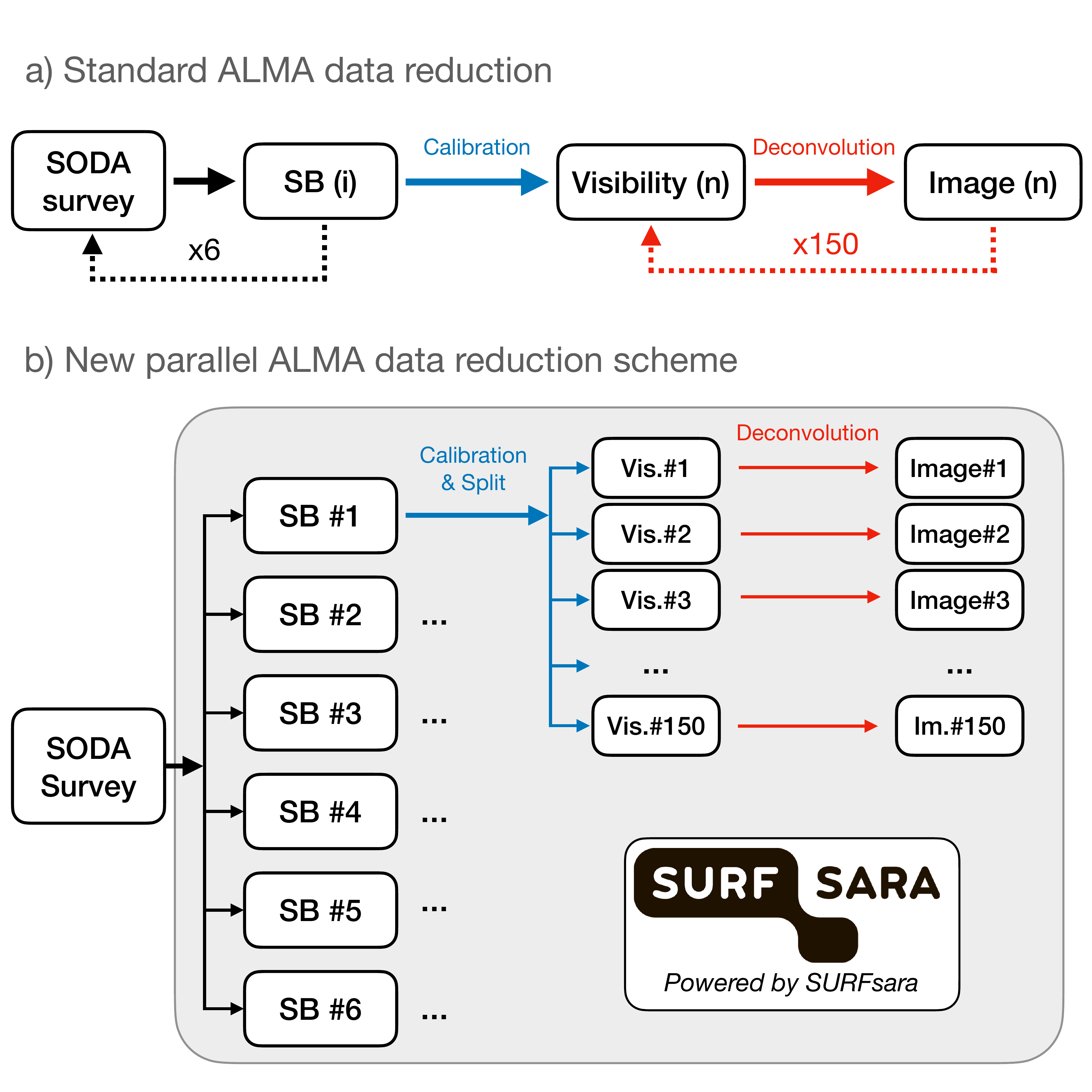}
     \caption{Data processing workflow for SODA. (a) Standard data reduction. (b) Parallel ALMA data reduction scheme implemented on the SPIDER platform that is part of SURFsara's Data Processing facilities.}
     \label{fig:SURFsara}
\end{figure}

\subsection{Observations}
We observed the 873 SODA disks using ALMA (Project ID: 2019.1.01813.S, PI: S. van Terwisga) between November 21 and December 2 2019 as part of Cycle 7. All observations were carried out with the most compact 12m-array configuration (C43-1) with a minimum of 41 antennas and under average weather conditions with precipitated water vapor PWV$=1.5-3.0$\,mm. All of our targets were observed in Band 6 including 4 spectral windows centered at frequencies of 224, 226, 240, and 242\,GHz, respectively, each with total bandwidth of 1.8\,GHz and resolution of 31\,MHz. The primary beam for this configuration is $\sim 1.1''$, with a Maximum Recoverable Scale (MRS) of $10.9''$. This choice of frequencies and array setups maximize the detection of line-free continuum emission from point-like sources in the shortest integration time possible.

We observe each target of our sample with a single ALMA 12-m array pointing for a total integration time of approximately 30 sec on source. However, the large number of targets in our sample makes the simultaneous scheduling of our observations unpractical. As part of our observation strategy, we then divided our SODA sample in 6 scheduling blocks (SBs), each typically containing 150 targets selected by proximity in declination. To guarantee the homogeneity of our SODA sample, all sources within a single SB were observed simultaneously. Moreover, each SB was executed a minimum of two times over the observing run. Pointing, bandpass, and flux calibrations (J0725-0054) were carried out at the beginning of each SB. Regular atmospheric and phase calibrations (J0541-0541) were obtained every 20 min. 

\subsection{Parallel data reduction at the SURFsara Data Processing Facility}
Our project develops a new data reduction framework for large ALMA surveys using high throughput data processing facilities. In standard data reduction schemes, both the calibration and deconvolution of ALMA observations is typically carried out in a sequential manner using personal computers or small server machines. This linear and iterative procedure makes the data processing extremely time consuming and largely inefficient for datasets with a large number of targets, such as ours (see Fig.\ref{fig:SURFsara}a). Instead, we introduce a new parallel scheme for data reduction at the SURFsara Data Processing facilities in Amsterdam (Fig.\ref{fig:SURFsara}b)\footnote{Since Jan. 1st 2021, SURFsara has become SURF:\url{https://www.surf.nl}}
\footnote{SODA is a pilot project of the new  “Advance ALMA data reduction using SPIDER” (PI: A. Hacar) carried out at at SURFsara by the Allegro Dutch ARC node in Leiden.
The new SODA technical developments are part of the new EMERGE project: \url{emerge.alvarohacar.com}}.

The original design and observing strategy of the SODA survey allow a direct parallelization of both calibration and imaging steps at the new SPIDER Data Processing Platform at SURFsara \footnote{\url{https://userinfo.surfsara.nl/documentation/data-processing-spider}; \url{https://spiderdocs.readthedocs.io/en/latest}}.
The new SODA reduction workflow is described in Figure \ref{fig:SURFsara}b. 
First, each of the 6 SB are independently calibrated using facility-provided scripts in order to recreate their corresponding visibilities. Second, each target field is extracted from the visibility file and individually imaged using identical cleaning parameters (see below). Each imaging process is run using an independent realization of CASA v5.6.1 and distributed over multiple cores and nodes by a Grid-based processing environment. We run independent CASA executions to calibrate all our 6 SBs. More importantly, we employ this new distributed scheme to execute up to 150 deconvolutions per SB simultaneously. Following the results of different performance tests, each of these 150 jobs is executed independently using 3 cores, which provide a total of 24 GB of memory per job, in order to optimize the CPU usage, as well as reduction and queuing times. The combination of high performance, flexibility, and number of cores at SURFsara-Spider allows us to reduce the total reduction time by 2 orders of magnitude ($\sim 6 \times 150 = 900$ times) compared to standard linear procedures. Independent experiments show that the use of CASA in parallel mode (i.e., mpicasa) does not introduce additional improvements in the SURFsara architecture. Therefore this method is not considered for our data reduction pipeline.

Originally developed for the reduction of single-field continuum data in SODA, this procedure will be implemented for the analysis of mosaics and spectral cubes in the near future (Hacar et al., in prep.). The reduction of the entire SODA sample required a total of $\sim$~ 3000 CPU hours. After optimization, our new parallel processing allows to run the full calibration and imaging processes of the entire SODA sample in less than a day. These data -- containing many independent pointings with relatively simple sources -- are particularly suited to this type of approach. A similar reduction strategy has been employed for the data reduction of large radio surveys obtained by interferometers such as the Low-Frequency Array (LOFAR)~\citep{mechev17,shimwell17} or Very Large Array (VLA) \citep{VLASSsurvey}. To our knowledge, this is the first time this type of parallel data processing is used in the case of ALMA observations.

The optimal use of parallel CASA executions requires the standardization of the imaging process.
For all our targets, we combine the four available continuum spectra windows in order to maximize the sensitivity of our maps, but did not further spectrally average them. We deconvolved each of our target fields with the CASA task {\it tclean} using standard Briggs weighting and a robust parameter of 0.5. Each of the resulting images is then corrected by their corresponding primary beam response. The individual maps show typical synthesized beam sizes of $\sim$~1.4''~$\times$~1.1'', that is, approximately 580 AU~$\times$~414 AU at the distance of Orion. 
The typical noise level of our maps is 0.08\,mJy\,beam$^{-1}$ which translates into a median sensitivity threshold of $1.5\,M_{\oplus}$ at the 4$\sigma$ level for our adopted opacities (see Section~\ref{sec:massassump}). This sensitivity is lower than that achieved in nearby low-mass star-forming regions, which achieve similar sensitivities of $0.2\,\textrm{M}_{\oplus}$~\citep[e.g.,][]{ansdell16,pascucci16}, but sufficient to detect the majority of sources in such regions assuming similar disk flux distributions. After imaging, we inspected the images for issues, but did not identify any. All our continuum maps are available as png figures and in FITS format on our webpage\footnote{\url{https://emerge.alvarohacar.com/results/soda}}. 

\subsection{Flux extraction}
To identify disks, we search within a 1.4'' radius around the catalog position for the brightest pixel in the reconstructed image. We use a $4\sigma$ detection limit, in order to ensure the number of false positive detections is $< 1$ for our dataset. This false positive rate includes both pure noise and the chance alignment of an IR excess source in the input catalog with a submillimeter galaxy in the background, following \citet{sadavoy19}. The assumption of unresolved flux is based on observations of nearby star-forming regions at higher resolution, which find median dust emission extents of $\sim 60$\,AU, while the largest objects are no larger than $\sim 250$\,AU, and exceedingly rare ($\lesssim 1\%$ of objects, e.g.~\citealp{vanterwisga18,long18,cieza19}). To test the validity of this hypothesis, we fit double Gaussians to the emission, using the location of the peak emission as a prior. We identify only 14 objects which have SEDs consistent with Class II objects and best-fit axis ratios inconsistent at the 4$\sigma$ level with the assumption of unresolved emission. These objects, and the results from the fitting procedure, are shown in Appendix A, Table~\ref{tab:app_fits} and Figure~\ref{fig:resolved_0}-\ref{fig:resolved_3}, and are marked out in Table~\ref{tab:results}. Additionally, we discuss two further sources with resolved emission but unresolved central sources in more detail in Section~\ref{sec:res}. Two types of objects are present: sources which seem to be well-fit by a double Gaussian, and which most likely represent objects with some contribution from (resolved) envelope emission, and five sources with residuals suggesting a compact (visual) multiple system ([MGM2012] 129, 754, 1000, 1090, and 1094). We have kept them in the catalog presented here to ensure consistency with regards to the source selection, and because of the small number of such objects. The impact of this flux extraction approach on multiple systems in particular is discussed in the following Section.

\section{Results}
\label{sec:res}
The SODA survey covers a large number of sources at low resolution. As a result, a wide range of stellar properties and environments are covered. In total, we detect 502 of 873 sources ($58 \pm 2.6 \%$) in the sample. Comparing this result to nearby low-mass SFRs, we find that the detection rate in SODA is in between Lupus ($66 \pm 6 \%$) and Cha I ($47 \pm 5 \%$), c.f. Fig.~\ref{fig:allclouds}. In comparison to the Class I sources in Orion presented by~\citet{tobin20}, which have detection rates $>85\%$ in a shallower survey, this is consistent with a continuous mass loss for the circumstellar material.

\begin{figure*}
\centering
   \includegraphics[width=17cm]{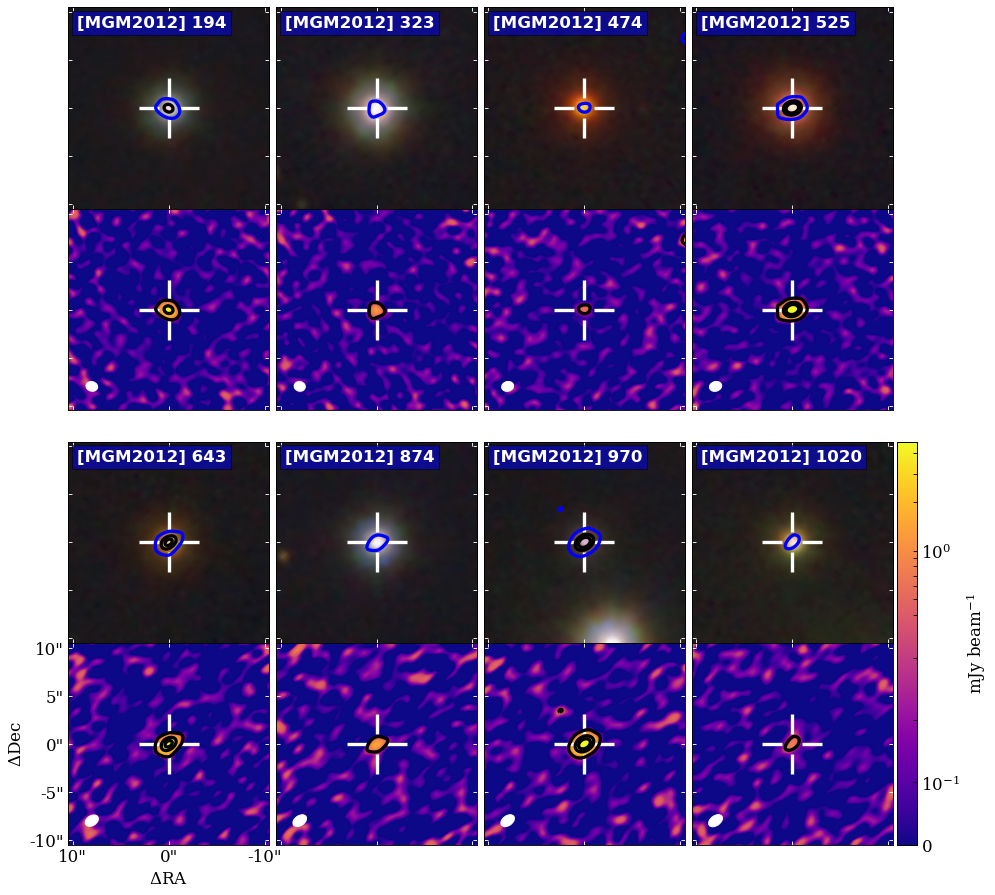}
     \caption{Images of selected millimeter-detected sources. For each source, the top panel shows near-infrared (NIR) data (J, H, and K$_s$-band) form the VISION survey~\citep{meingast16} in colors, and 5, 50, and 100$\sigma$ contour levels. The bottom panel shows the ALMA image at $225\,GHz$ only. White crosshairs mark the positions of detected sources. The top panels show sources with fluxes $> 4.8$\,mJy, bottom panels sources with fluxes $< 1.6$\,mJy. The beam is shown in white in the ALMA images, in the lower left.}
     \label{fig:panels}
\end{figure*}

\begin{figure*}
\centering
   \includegraphics[width=17cm]{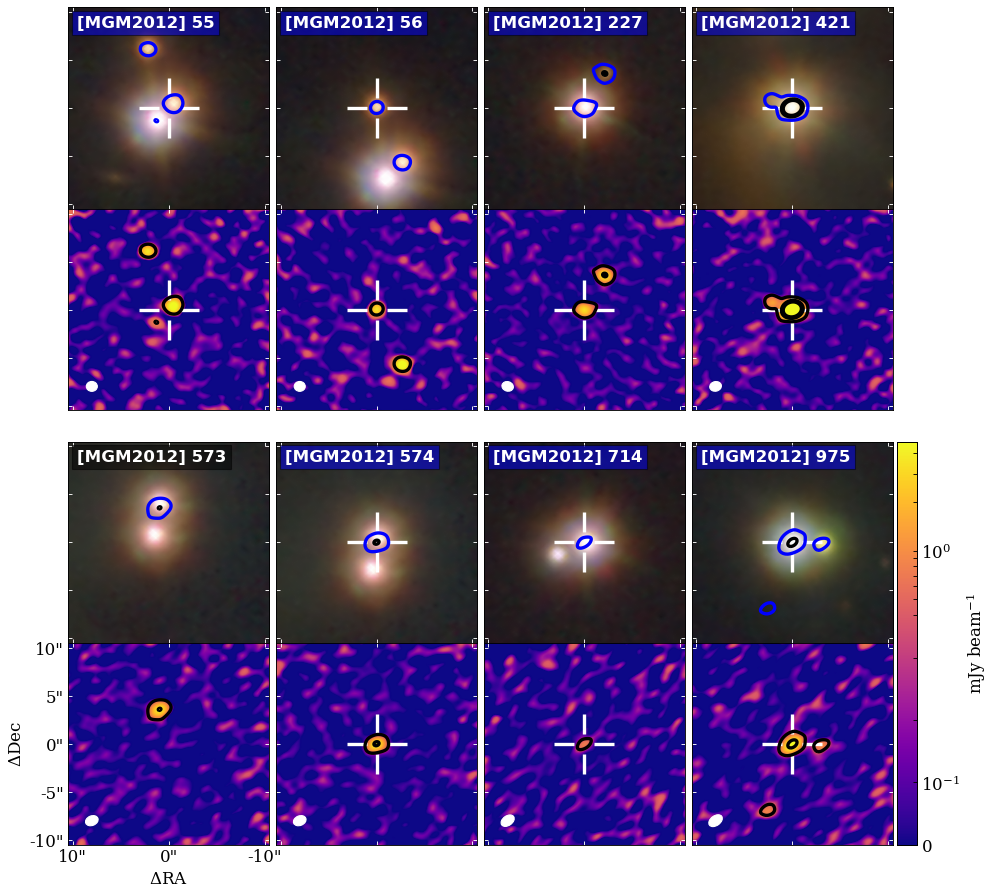}
     \caption{Images of selected (candidate) multiple systems. For each source, the top panel shows NIR data (J, H, and K$_s$-band) form the VISION survey~\citep{meingast16} in colors, and 5, 50, and 100$\sigma$ contour levels. The bottom panel shows the ALMA image at $225\,GHz$ only. White crosshairs mark the source positions of detected sources. Nondetections are marked with grayed-out names. The beam is shown in white in the ALMA images, in the lower left.}
     \label{fig:panels2}
\end{figure*}

Figure~\ref{fig:panels} shows a number of individual detections, both of millimeter-bright disks (in the top two rows) and fainter (but still detectable) disks, and their stellar hosts. The top panel shows, for each source, the ALMA contours on top of near-infrared (J, H, K$_s$) images from the Vienna Survey in Orion (VISION) using the VISTA telescope~\citep{meingast16}. This survey has a higher resolution than the input catalog, which is limited by the {\it Spitzer} beam. From the figure it is clear that the sample covers a wide range of disk masses. As previous work has shown, there is a correlation -- albeit with a large scatter -- between the masses of disks and their host stars~\citep[e.g.,][]{andrews13,ansdell16,pascucci16}. Thus, before comparing the SODA sample to other star-forming regions, we should confirm the similarity of the underlying Initial Mass Function (IMF) of L1641 and L1647. L1641 is deficient in B- and late-O-type stars~\citep{hsu12,hsu13}, but similar to other nearby SFRs at lower masses. This means the sample is truly representative of that in other regions, where such high-mass stars are also often absent, and M- and K-type stars dominate the stellar mass distribution.

Multiplicity may also affect the properties of disks. In Fig.~\ref{fig:panels2}, a selection of visual multiples with at least one ALMA continuum detection of a disk are shown. Unfortunately, multiplicity is not a well-constrained quantity for the SODA survey sample. Both the input catalog resolution of $1''$ and the resolution of our ALMA data prevent us from resolving the most compact binaries (with separations $\lesssim 400$\,AU). For such objects, we effectively consider the system's average continuum luminosity. The cumulative distribution of total system masses for the large sample of multiple systems in nearby SFRs presented in~\citet{zurlo20} lacks massive ($> 50\,M_{\oplus}$) systems, but the mass distributions are otherwise identical. Thus, we do not expect the way we treat these systems and the marginally resolved ones shown in Appendix A to affect our estimates of median disk luminosities and masses.

In cases where apparent (visual) companions are present, at separations $< 10''$ (4000 AU), we have included both sources if both sources have Spitzer photometry indicating a circumstellar disk, as in [MGM2012] 55 and 56 and [MGM2012] 573 and 574 in Figure~\ref{fig:panels2}. Alternatively, if only one source is included, we only include that source. This is for instance the case for [MGM2012] 227 and 975, which are well-resolved, and for the marginally resolved [MGM2012] 129 and 421. If we assume that the trends of disk properties with multiplicity as demonstrated by~\citet{harris12} and~\citet{akeson19} in Taurus, and~\citet{zurlo20,zurlo21} in Lupus and Ophiuchus also apply in Orion, we do not expect this approach to lead to a (strong) bias. The primary star's disk in a wide binary ($>300$\,AU in ~\citealp{harris12}, $> 140$\,AU in~\citealp{akeson19}) has a similar luminosity distribution to the disks in single Class II YSOs. Due to the resolution of our data, this separation criterion is always met for (marginally) resolved systems. While a detailed confirmation is beyond the scope of this article, we may therefore expect no large bias will arise from the approach used here. Moreover, not all visual multiple systems are true multiples: they may merely be chance alignments. For objects like [MGM2012] 714's companion candidate, which does not have an IR excess, adding nondetections for the visual companions leads to a more inconsistent and biased catalog.

\begin{figure*}
\centering
   \includegraphics[width=17cm]{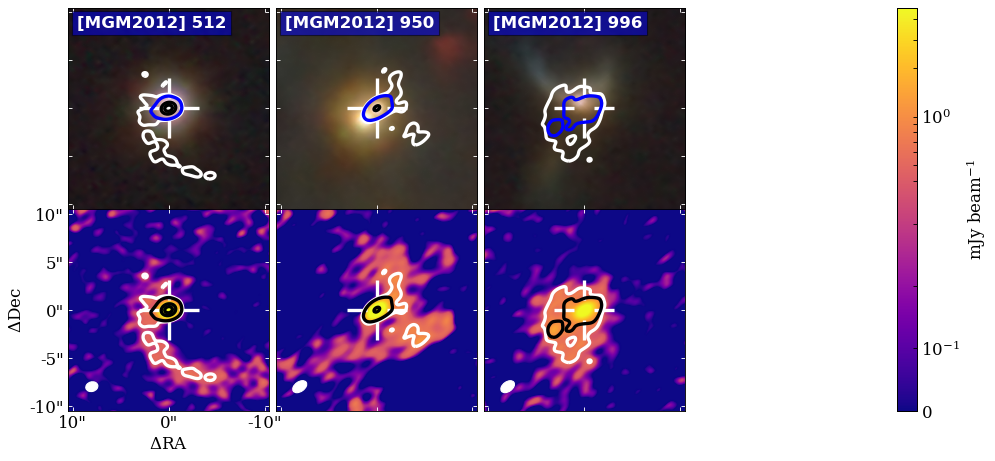}
    \caption{Images of YSOs associated with extended millimeter emission. For each source, the top panel shows NIR data (J, H, and K$_s$-band) form the VISION survey~\citep{meingast16} in colors, and 3, 5, 50, and 100$\sigma$ contour levels. The bottom panel shows the ALMA image at $225\,GHz$ only. White crosshairs mark the source positions of detected sources. The beam is shown in white in the ALMA images, in the lower left.}
     \label{fig:weirdpanels}
\end{figure*}

Apart from unresolved disks, we also identify two Class II sources associated with extended millimeter-continuum emission. Fig.~\ref{fig:weirdpanels} shows these objects: [MGM2012] 512, which was previously identified by~\citet{grant21}, and [MGM2012] 950, where to our knowledge no extended emission has previously been identified. Both of these sources show striking and, unexpectedly, resolved continuum emission. [MGM2012] 512 is particularly noteworthy, and appears to have an arc of dust close to the star. Such extended emission is not expected to be found in Class II sources, which are expected to have lost their envelopes, although they have counterparts in younger, more embedded objects~\citep{pineda20}. As~\citet{grant21} discuss, however, this could be due to either a late accretion event, a perturbed disk (although no suitable candidates can be identified in either 2MASS or VISION), or a younger, still partially embedded protostar viewed from nearly face-on. Face-on embedded protostars are indeed expected to have much bluer spectral indices than those seen at higher inclinations~\citep{whitney03,crapsi08}. As~\citet{grant21} discuss, the classification of the source is ambiguous in the literature, and~\citet{carattiogaratti12} consider it to be a Class I object. On the other hand, the late accretion scenario has also been proposed to explain the scattered-light and submillimeter molecular line observations of an arc of material accreting onto the Class II SU Aur disk~\citep{ginski21}.

A misclassification is the most likely explanation for the nature of the [MGM2012] 950 source: to the northwest of the source, there is tentative evidence of an outflow-like structure in the VISION data. However, the Class I [MGM2012] 996 source, which we observed in this survey (but do not include in our statistical analysis due to its classification) has extended continuum emission but a quite different, more centrally concentrated morphology at the same resolution.

Regardless of the origin of the observed extended dust tails seen in these two sources, the advantage of the size of this homogeneous dataset is that it is possible to say that this kind of morphology must be rare, occurring only $0.23 \pm 0.16 \%$ of the time. Thus, if these structures are the result of younger sources contaminating the sample, this demonstrates that such contamination is not common. This is further supported by the small number of objects with possible resolved emission in Appendix A: the most likely cause for this extended emission is the presence of a (remnant) envelope. In either case, these objects with significant large-scale continuum emission are excellent candidates for follow-up studies of how circumstellar material accretes onto the star and disk, both at millimeter wavelengths and in scattered light.

\label{sec:results}
\subsection{Disk luminosities and masses}
In this subsection, we discuss how we inferred the millimeter luminosities and masses from the fluxes measured with our survey, as listed in Table~\ref{tab:results}. 
Generally, disk masses are inferred by assuming that their millimeter continuum flux is optically thin, with a known opacity and distance. In this case, the mass is found from:

\begin{equation}
\label{eq:mass}
	M_{\rm{dust}} = \frac{F_{\nu} d^2}{\kappa_{\nu} B_{\nu}(T_{\rm{dust, eff}})}
\end{equation}

We first discuss the relevant terms and assumptions in this equation. Inferring the disk luminosities only requires assuming one parameter, the distance, which is discussed in Section~\ref{sec:distances}. For disk masses, several additional assumptions must be made. In Section~\ref{sec:massassump} we detail our choices in this regard, and how they may impact the comparison with other surveys in the rest of this article.
	
	\subsubsection{Temperature and opacity assumptions for disk masses}
	\label{sec:massassump}
	For this survey, we assume $T_{\rm{dust, eff}} = 20$\,K. This is based on the result in~\citet{andrews05} for disks in Taurus, and followed by most other disk surveys. However, this temperature implicitly carries with it assumptions on disk radius: more compact disks will have higher effective temperatures. Likewise, large radiation fields (such as in the ONC) or simply hotter host stars will lead to intrinsically hotter disks. However, this survey's sample, located far ($\geq 4$\,pc) away from the massive stars in the Trapezium, should not be truncated by external photo-evaporation, nor be intrinsically hotter than those of other SFRs. Even assuming a higher temperature of for instance $30$\,K will, however, not have a large effect on the measured dust mass.
	
	Likewise, the assumptions of optically thin emission and the opacity $\kappa_{\nu}$ assumed in the calculation of disk masses are a source of much unquantified uncertainty. There is now increasing evidence that part of the disk can be optically thick even at the wavelengths used for this survey (e.g.,  \citealp{andrews05,tripathi17,zhu19,macias21}, the latter of which find optically thick emission even at 2.1\,mm). The net effect of this is that we possibly underestimate our disk masses to some (model-dependent) extent. 
	
	Opacities for the millimeter-sized grains can also vary. Here, we used common assumptions on the opacities of the grains, derived from~\citet{beckwith90}: $\kappa_{\nu = 230\,\rm{GHz}} = 2.3$\,cm$^2$\,g$^{-1}$, and a power-law frequency-dependence with $\beta = 1$. In the recent literature, these are commonly-used values~\citep[e.g.,][]{ansdell16,ansdell17,ansdell20,vanterwisga19b,vanterwisga20}, but~\citet{barenfeld16} and~\citet{pascucci16} both used $\beta = 0.4$ for Band 7 observations. The effect of this, however, is small, especially in the light of other existing uncertainties on disk temperatures, radii, and optical depths; see also~\citet{vanterwisga19b}.
	
\begin{table*}
\caption{Continuum fluxes and masses for the SODA sample. The full table is available in electronic form online.}
\label{tab:results}
\centering
\begin{tabular}{cccccc}\hline\hline
[MGM2012] & RA & Dec & $d$ & F$_{225\,\textrm{GHz}}$ & $M_{\textrm{dust}}$ \\
 & [J2000] & [J2000] & [pc] & [mJy] & [M$_{\oplus}$] \\
\hline
2 & 05:42:30.53 & -10:10:48.62 & 454 & $ < 0.4$ & $ < 2.4$ \\
3 & 05:43:01.59 & -10:07:50.40 & 454 & 0.38 $\pm$ 0.092 & 2.4 $\pm$ 0.58 \\
5 & 05:42:31.78 & -10:05:26.56 & 454 & $ < 0.3$ & $ < 2.0$ \\
6 & 05:42:37.98 & -10:03:42.94 & 454 & $ < 0.3$ & $ < 1.9$ \\
7 & 05:42:59.94 & -10:03:40.62 & 454 & 0.49 $\pm$ 0.088 & 3.1 $\pm$ 0.56 \\
8 & 05:43:00.00 & -10:03:36.06 & 454 & 0.85 $\pm$ 0.100 & 5.4 $\pm$ 0.63 \\
9 & 05:42:57.87 & -10:03:36.34 & 454 & $ < 0.3$ & $ < 2.2$ \\
10 & 05:42:37.07 & -10:03:30.00 & 454 & $ < 0.4$ & $ < 2.4$ \\
11 & 05:42:44.45 & -10:03:18.78 & 454 & 4.09 $\pm$ 0.097 & 25.8 $\pm$ 0.61 \\
12 & 05:42:52.84 & -10:02:42.66 & 454 & $ < 0.4$ & $ < 2.3$ \\
13 & 05:42:52.58 & -10:02:29.02 & 454 & $ < 0.3$ & $ < 2.2$ \\
15 & 05:42:12.35 & -10:02:10.24 & 454 & $ < 0.4$ & $ < 2.3$ \\
16 & 05:42:34.32 & -10:01:52.30 & 454 & $ < 0.4$ & $ < 2.3$ \\
17 & 05:42:27.76 & -10:01:51.40 & 454 & 10.11 $\pm$ 0.084 & 63.8 $\pm$ 0.53 \\
18 & 05:42:25.72 & -10:01:48.28 & 454 & $ < 0.4$ & $ < 2.5$ \\
20 & 05:42:34.89 & -10:01:46.62 & 454 & 17.49 $\pm$ 0.107 & 110.4 $\pm$ 0.67 \\
21 & 05:42:28.57 & -10:01:42.48 & 454 & $ < 0.3$ & $ < 2.1$ \\
22 & 05:42:31.53 & -10:01:36.82 & 454 & 0.96 $\pm$ 0.078 & 6.0 $\pm$ 0.49 \\
25 & 05:42:00.07 & -10:01:11.76 & 454 & $ < 0.3$ & $ < 2.1$ \\
26 & 05:42:38.18 & -10:01:08.40 & 454 & $ < 0.3$ & $ < 2.2$ \\
27 & 05:42:41.54 & -10:01:06.02 & 454 & $ < 0.4$ & $ < 2.2$ \\
28 & 05:43:05.05 & -10:01:02.70 & 454 & 0.92 $\pm$ 0.091 & 5.8 $\pm$ 0.58 \\
29 & 05:42:36.48 & -10:01:03.16 & 454 & 0.38 $\pm$ 0.087 & 2.4 $\pm$ 0.55 \\
30 & 05:42:15.38 & -10:00:58.78 & 454 & $ < 0.4$ & $ < 2.4$ \\
33 & 05:42:15.97 & -10:00:41.08 & 454 & 0.40 $\pm$ 0.087 & 2.5 $\pm$ 0.55 \\
34 & 05:42:30.49 & -10:00:38.62 & 454 & $ < 0.4$ & $ < 2.3$ \\
35 & 05:42:24.32 & -10:00:36.18 & 454 & $ < 0.4$ & $ < 2.6$ \\
36 & 05:42:39.10 & -10:00:33.86 & 454 & 1.19 $\pm$ 0.091 & 7.5 $\pm$ 0.57 \\
38 & 05:43:04.90 & -10:00:20.30 & 454 & 1.30 $\pm$ 0.092 & 8.2 $\pm$ 0.58 \\
39 & 05:42:32.34 & -10:00:15.96 & 454 & $ < 0.4$ & $ < 2.3$ \\
41 & 05:42:34.26 & -09:59:49.08 & 454 & $ < 0.3$ & $ < 2.2$ \\
43 & 05:42:16.90 & -09:59:42.34 & 454 & 14.53 $\pm$ 0.119 & 91.7 $\pm$ 0.75 \\
45 & 05:42:37.04 & -09:59:37.82 & 454 & 0.67 $\pm$ 0.087 & 4.2 $\pm$ 0.55 \\
46 & 05:42:36.50 & -09:59:12.92 & 454 & $ < 0.4$ & $ < 2.4$ \\
47 & 05:42:35.84 & -09:58:55.32 & 454 & 2.39 $\pm$ 0.105 & 15.1 $\pm$ 0.67 \\
48 & 05:42:42.63 & -09:58:37.72 & 454 & $ < 0.4$ & $ < 2.4$ \\
49 & 05:42:10.39 & -09:58:38.22 & 454 & 22.41 $\pm$ 0.104 & 141.4 $\pm$ 0.66 \\
50 & 05:42:57.57 & -09:58:22.62 & 454 & $ < 0.3$ & $ < 2.1$ \\
51 & 05:42:31.65 & -09:57:41.40 & 454 & 0.66 $\pm$ 0.089 & 4.1 $\pm$ 0.56 \\
52 & 05:42:51.75 & -09:57:41.12 & 454 & 13.79 $\pm$ 0.108 & 87.0 $\pm$ 0.68 \\
\hline
\end{tabular}
\tablefoot{
\tablefoottext{a}{Source fit indicative of possible extended emission.}
}\end{table*}

\begin{table*}
\caption{Continuum fluxes for Flat spectrum, Class I and 0 sources in the SODA sample.}
\label{tab:embedded}
\centering
\begin{tabular}{ccccccl}\hline\hline
[MGM2012] & HOPS & RA & Dec & d & F$_{225\,\textrm{GHz}}$ & Class \\
 & & [J2000] & [J2000] & [pc] & [mJy] &  \\[3pt]
\hline
354 & 256 & 05:40:45.27 & -08:06:42.14 & 404 & 1.23 $\pm$ 0.199 & 0 \\
518 & 288 & 05:39:56.00 & -07:30:27.60 & 401 & 417.95 $\pm$ 8.853 & 0 \\
895 & 173 & 05:36:26.05 & -06:25:05.22 & 384 & $ < 8.7$ & 0 \\
896 & 380 & 05:36:25.34 & -06:25:02.44 & 384 & $ < 17.0$ & 0 \\
\hline
42 &  & 05:42:17.83 & -09:59:48.00 & 454 & 13.88 $\pm$ 3.229 & I \\
44 &  & 05:42:17.50 & -09:59:40.66 & 454 & 20.61 $\pm$ 0.695 & I \\
69 &  & 05:42:56.29 & -09:50:51.80 & 454 & 24.10 $\pm$ 0.686 & I \\
213\tablefootmark{a} & 209 & 05:42:52.92 & -08:41:41.18 & 431 & 0.45 $\pm$ 0.132 & I \\
257 & 228 & 05:41:34.19 & -08:35:27.90 & 421 & 29.02 $\pm$ 0.444 & I \\
423 & 273 & 05:40:20.91 & -07:56:24.60 & 404 & 61.33 $\pm$ 2.659 & I \\
519 & & 05:39:56.92 & -07:30:23.46 & 401 & $ < 21.8$ & I \\
524 & 289 & 05:39:56.73 & -07:30:05.84 & 401 & $ < 13.8$ & I \\
537 & 287 & 05:40:08.79 & -07:27:27.94 & 401 & 36.25 $\pm$ 4.000 & I \\
627 & 135 & 05:38:45.33 & -07:10:56.08 & 388 & 18.25 $\pm$ 4.373 & I \\
649\tablefootmark{a} & 136 & 05:38:46.53 & -07:05:37.38 & 387 & 22.63 $\pm$ 0.168 & I \\
678 & 140 & 05:38:46.26 & -07:01:53.72 & 386 & 19.04 $\pm$ 0.431 & I \\
709 & 148 & 05:38:39.59 & -06:59:30.70 & 386 & $ < 16.0$ & I \\
879 & 188 & 05:35:29.85 & -06:26:58.20 & 384 & 269.02 $\pm$ 3.290 & I \\
933 & 178 & 05:36:24.62 & -06:22:41.14 & 384 & 20.15 $\pm$ 9.283 & I \\
940 & & 05:36:21.16 & -06:22:42.60 & 384 & $ < 742.8$ & I \\
944 & 181 & 05:36:19.50 & -06:22:11.94 & 384 & 3.33 $\pm$ 0.768 & I \\
996\tablefootmark{a} & 185 & 05:36:36.97 & -06:14:58.80 & 384 & 8.63 $\pm$ 0.340 & I \\
\hline
197 & 207 & 05:42:38.59 & -08:50:18.84 & 434 & $ < 2.5$ & Flat \\
246 & 211 & 05:42:58.34 & -08:37:44.24 & 429 & 1.68 $\pm$ 0.185 & Flat \\
298\tablefootmark{a} & 260 & 05:40:19.40 & -08:14:16.28 & 405 & 28.76 $\pm$ 0.160 & Flat \\
301 & 259 & 05:40:20.90 & -08:13:55.24 & 405 & 3.76 $\pm$ 0.563 & Flat \\
315\tablefootmark{a} & 242 & 05:40:48.54 & -08:11:08.94 & 405 & 4.81 $\pm$ 0.137 & Flat \\
358 & 252 & 05:40:49.92 & -08:06:08.36 & 404 & 2.28 $\pm$ 0.407 & Flat \\
369\tablefootmark{a} & 255 & 05:40:50.58 & -08:05:48.70 & 404 & 10.96 $\pm$ 0.154 & Flat \\
390\tablefootmark{a} & 284 & 05:38:51.50 & -08:01:27.42 & 403 & 8.13 $\pm$ 0.133 & Flat \\
417 & & 05:41:21.05 & -07:57:07.44 & 404 & $ < 4.4$ & Flat \\
550 & 120 & 05:39:34.35 & -07:26:11.28 & 398 & $ < 14.5$ & Flat \\
587 & 128 & 05:38:52.02 & -07:21:05.88 & 394 & 5.67 $\pm$ 0.256 & Flat \\
618\tablefootmark{a} & 134 & 05:38:42.79 & -07:12:43.88 & 389 & 16.17 $\pm$ 0.164 & Flat \\
624\tablefootmark{a} & 132 & 05:39:05.37 & -07:11:05.16 & 390 & 3.33 $\pm$ 0.184 & Flat \\
696 & 141 & 05:38:48.00 & -07:00:48.24 & 386 & $ < 8.6$ & Flat \\
721 & 149 & 05:38:40.50 & -06:58:22.02 & 386 & 106.60 $\pm$ 1.433 & Flat \\
897 & 174 & 05:36:25.87 & -06:24:58.66 & 384 & 30.88 $\pm$ 0.616 & Flat \\
919 & 179 05:36:21.88 & -06:23:30.00 & 384 & 41.58 $\pm$ 3.528 & Flat \\
\hline
\end{tabular}
\tablefoot{
\tablefoottext{a}{Source reclassified based on~\citet{furlan16}.}
}
\end{table*}

	\subsubsection{The distance structure of Orion A}
	\label{sec:distances}
	The assumed distance has a potentially important effect on the conversion between fluxes and masses in Eq.~\ref{eq:mass}, and not only between SFRs. Given the size of the Orion A cloud, which spans multiple parsecs projected on the sky, we must be careful to take into account the three-dimensional structure of both the cloud and its young stellar population as best as possible, and literature results suggest that this is indeed quite complex~\citep{kounkel17,kounkel18,grossschedl18,zari19}. In particular,~\citet{grossschedl18} find quite pronounced distance differences between the southernmost and central parts of Orion A, of some 80\,pc.
	
	With the advent of the {\it Gaia} satellite, the distances and proper motions of nearby stars are better-known than ever before. However, the generally high optical extinctions mean that not all Class II stars in our sample have high-quality parallaxes in DR2 or EDR3. To solve this problem, we use the same approach as~\citet{grossschedl18}: YSOs with well-known distances from DR2 are placed in 16 bins in galactic longitude (oversampled by a factor of two), and the distances of the YSOs in our sample are interpolated based on the distances of these bins. After correcting for the distances structure, we are left with a residual uncertainty in the distance of the sources of $\sim 5\%$ (20 pc), with small variations in galactic longitude. This reflects both the uncertainty in the {\it Gaia} data and the intrinsic three-dimensional structure of the cloud.

	\subsection{Disk mass distribution in SODA}
\begin{figure*}[ht]
\centering
    \includegraphics[width=17cm]{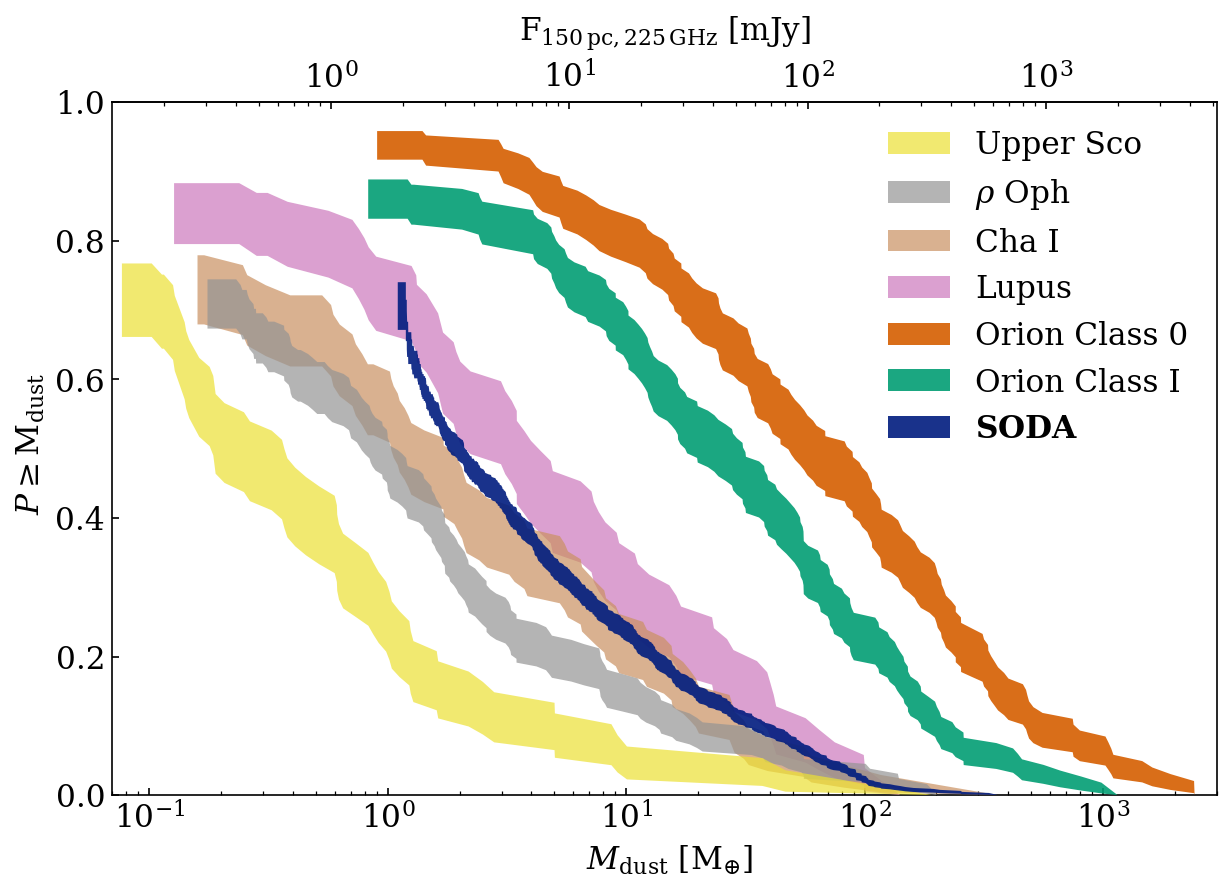}
     \caption{Disk mass distributions inferred with the Kaplan-Meier estimator in Orion A and several nearby star-forming regions. SODA (dark blue) versus Class I and Class 0 disks in Orion \citep{tobin20} (orange and green), and disks in Lupus \citep{ansdell16} (purple), Chamaeleon I \citep{pascucci16} (brown), $\rho$ Oph \citep{williams19} (gray), and Upper Sco \citep{barenfeld16} (yellow).}
     \label{fig:massdist_SODA_all}
\end{figure*}

Figure~\ref{fig:massdist_SODA_all} shows the inferred disk mass distribution using the Kaplan-Meier estimator for the full SODA Class II sample of 873 disks. We find a median disk dust mass of $1.97 ^{+0.15}_{-0.19} \,M_{\oplus}$. This figure also compares the observed disk mass distribution to several well-characterized populations of Class II disks in nearby star-forming regions, as well as to the Class I and Class 0 disk mass distributions in Orion A and B combined. Throughout this section and in the following, we use the \texttt{lifelines} Python package~\citep{lifelines} for fitting the (censored) disk mass distributions. For log-rank correlation tests, we follow~\citet{feigelson85} in subtracting our data from a constant, since \texttt{lifelines} only implements this test for right-censored data.

 The large sample size allows us to estimate the properties of this distribution with unprecedented accuracy. Errors in the probability distribution are on average $\pm 0.017$, compared to for instance $\pm 0.05$ for Taurus and Lupus, an improvement of a factor of 3. Moreover, it is noteworthy how well-sampled even the upper end of the disk mass range is: SODA characterizes even the occurrence of the most massive ($M_{{\rm disk}} > 100 M_{\oplus}$) well, despite their rarity; see also Sect.~\ref{sec:mostmassive}.

	Typically, in the literature, protoplanetary disk mass distributions are described using the Kaplan-Meier estimator, which is a powerful, nonparametric method for describing censored random variables. However, it is -- formally -- only useful if the censoring is not dependent on the variable in question. In these types of surveys, that is the case, leading to potential biases. These biases will primarily influence the lower end of the disk mass range probed by a survey, where incompleteness is the largest. Since our survey depth is similar to that of other works, we conclude that our results should be comparable to those of other surveys to approximately the same masses.

	Instead of using nonparametric descriptions of the disk mass distribution, however, we may also -- if the nature of the distribution is known a priori -- use parametric methods to compare regions. As~\citet{williams19} showed, many protoplanetary disk mass distributions appear to follow log-normal distributions, with highly similar standard deviations, but potentially quite different median disk masses. If we can fit log-normal distributions to our observed disk populations, the mass-dependent censoring problem is less relevant. This is particularly important for samples with fewer sources, as well as for the faint tail of the disk mass distribution. Moreover, fitting a log-normal distribution allows us to directly compare (including statistical errors) the median disk masses of different regions. However, it is important to test if this parametric distribution is indeed a good description of the property we study here.
	
\begin{figure}
	\resizebox{\hsize}{!}{\includegraphics{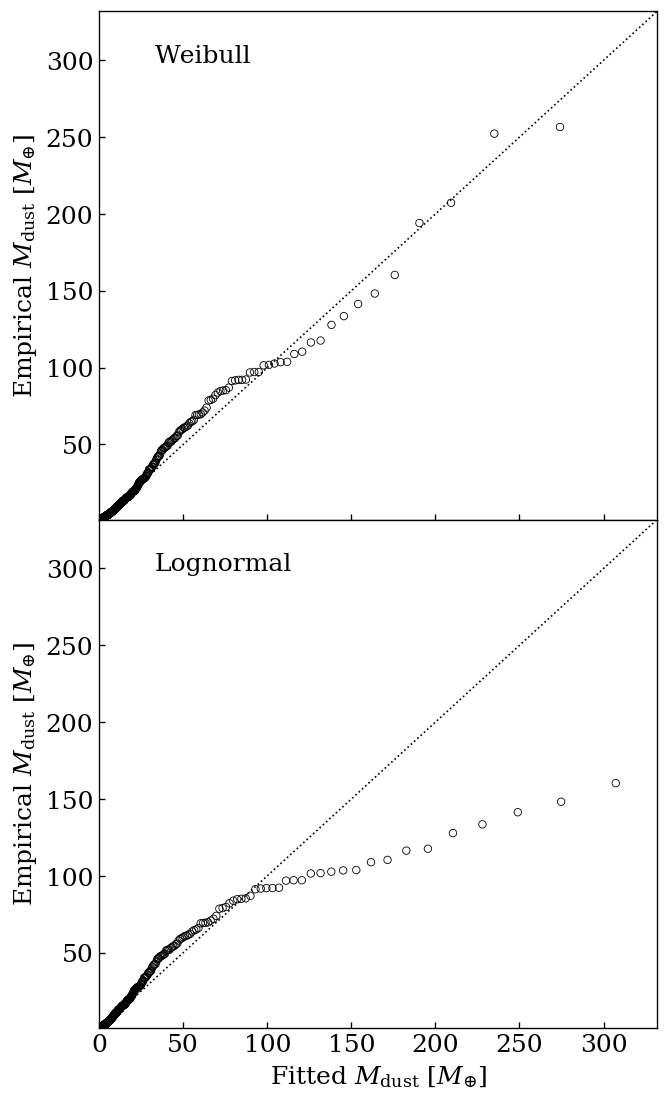}}
    \caption{Quantile-quantile plots of the empirical distribution (inferred with a Kaplan-Meier estimator) and fitted quantiles using a Weibull distribution (top panel) and log-normal distribution (bottom panel).}
     \label{fig:qqplot}
\end{figure}

	Interestingly, Figure~\ref{fig:qqplot} shows that SODA sample is not best described by a log-normal distribution, but by a Weibull distribution. Like the log-normal distribution, this distribution has two parameters; its cumulative distribution function is $1-\exp{-(x/\lambda)^k}$. That the Weibull distribution is indeed a better fit is confirmed from a comparison of the Akaike Information Criterion (AIC): $\Delta_{\rm{AIC}} = 16$ in favor of the Weibull distribution. Figure~\ref{fig:qqplot} shows the quantile-quantile plots of the empirical distribution of source masses versus a Weibull (top panel) and log-normal (bottom panel) distribution fit to the data. In this figure the axes are linear, not logarithmic, in disk mass. It is clear that the difference is driven by the better fit of the Weibull distribution at masses $> 100\,\textrm{M}_{\oplus}$: the SODA sample has a more massive tail than a log-normal distribution can account for. However, the values for the median dust mass agree to within their respective errors: these are $2.2 \pm 0.19\,\rm{M}_{\oplus}$ for the Weibull, and $2.2^{+0.17}_{-0.17}$ for the log-normal distribution. The differences between the information criteria for these two-parameter distributions are also small. Indeed, neither distribution perfectly describes the intermediate-mass regime: there is more complexity hiding in these data. 

	Similar, but much more extreme, behavior is seen for the disk masses in the $\sigma$ Ori cluster~\citep{ansdell17}. The authors find a steep disk mass gradient in space in that region. As a result the disk mass distribution is likewise not well described by a single log-normal disk mass distribution, but shows an excess of massive sources. This suggests that inhomogeneity in the underlying disk sample may drive the observed behavior away from a single log-normal distribution in the SODA sample as well. In Section~\ref{sec:disc}, we investigate possible drivers of this behavior in the SODA sample in greater detail.

\subsection{Comparison to literature results}
The sample sizes in SODA allow us to split up the sample by the subclouds in Orion A, L1641 and L1647, respectively. Similar data to those presented here (in terms of resolution and frequency) exist for a subset of the SODA sample in L1641, by~\citet{grant21}. That sample was based on Class II objects detected with {\it Herschel} in the HOPS fields~\citep{furlan16}. Surprisingly, the disk mass distribution inferred by these authors has a very high median mass, even including nondetections, which they consider to be the likely consequence of a selection bias. In the most pessimistic case, considering all nondetections in {\it Herschel}-bands to be nondetections with ALMA, the disk mass distribution for L1641 could be similar to that of Upper Sco, with a very low detection rate. However, from this sample alone the bias cannot be accurately determined.

\begin{figure}
	\resizebox{\hsize}{!}{\includegraphics{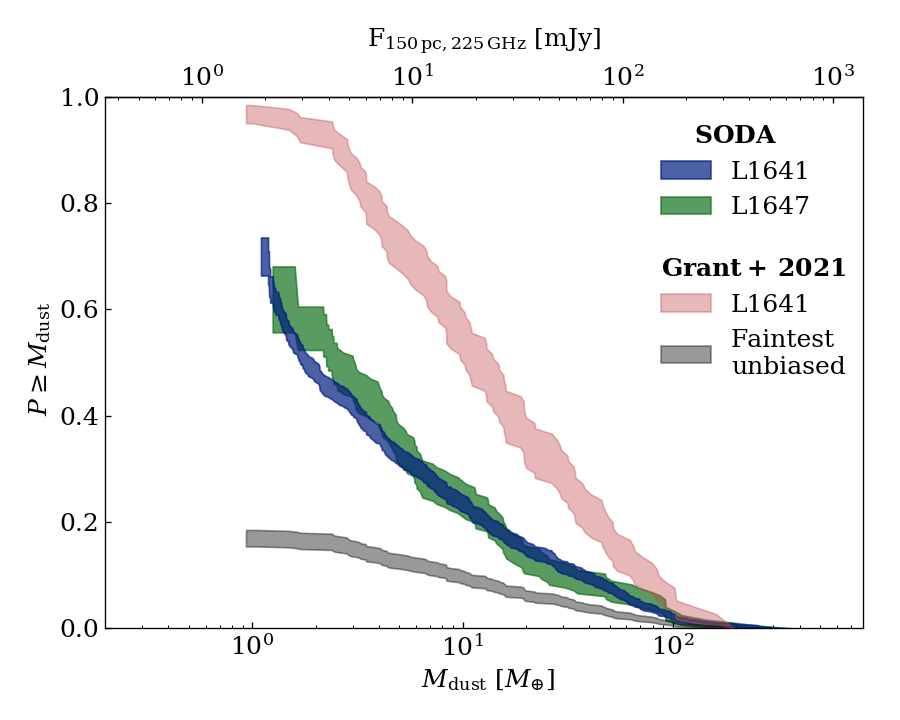}}
    \caption{Comparison of SODA disk masses to previous literature. Disk mass (or luminosity) distributions in L1641 (blue) and L1647 (green) from SODA, as well as the observed disk mass distribution in the sample from~\citet{grant21} (pink) and a scenario in which all of the unobserved sources in~\citet{grant21} are also nondetections at their sensitivity (gray).}
    \label{fig:G21}
\end{figure}

In Figure~\ref{fig:G21}, the SODA data are split up by cloud and compared to the observed sample from~\citet{grant21}. Two results are readily apparent (and confirmed statistically, using a cutoff of $p = 0.05$ in a log-rank test): L1641 and L1647 have highly similar disk mass distributions, and the {\it Herschel}-detected subsample in L1641 (shown in pink) is indeed biased toward higher disk masses. At the same time, it is clear that a significant fraction of sources not detected by {\it Herschel} is still detectable by ALMA at the effective sensitivity of these observations, as shown by the difference between the gray distribution and the SODA disk mass distributions.

\subsection{The most massive disks in Orion A}
\label{sec:mostmassive}
The most massive protoplanetary disks in nearby star-forming regions have been a crucial population in the study of disk evolution. These objects are the easiest to observe at high resolution, and may correspond to stars with optically thick material at the largest radii~\citep{andrews18}. As such, they allow us to cast light on the formation of, potentially, distant planets, and planetary system architectures that are difficult to observe in main-sequence systems. These massive disks are also relevant since they are the most plausible systems for identifying signatures of gravitational instability. This is especially the case if they are radially extended as well as massive (which is suggested by the disk luminosity-radius relation, see e.g. \citealt{andrews18b}). Such instabilities have now been suggested to occur in a handful of massive nearby disks~\citep[e.g.,][]{huang18,booth20}. As such, it is a sample that is also particularly interesting for follow-up observations with ALMA.

Figure~\ref{fig:pointingmap} shows the locations of all disks with masses (assuming optically thin emission) higher than $100\,M_{\oplus}$ in dust in Orion A. We find 20 such objects, larger than any other such sample, and the first complete sample of massive protoplanetary disks located within the same star-forming region. The most massive object is [MGM2012]-540, with $M_{\textrm{dust}} = 364 \pm 0.5\,M_{\oplus}$. In L1641,~\citet{grant21} previously identified only 6 disks with masses more than $100\,M_{\oplus}$; this survey increases that number to 16 in the same part of the cloud, emphasizing the importance of an unbiased survey for characterizing the upper end of the disk mass distribution.

No clear pattern is visible in their occurrence across the cloud, with the possible exception of a smaller fraction of such sources in L1641-C and L1641-N. Interestingly, these sources also seem to be found only where the cloud is densest, suggesting that they may be younger on average than the full disk sample. However, the relatively small number ($2\%$) of such sources means that it is difficult to identify significant patterns in their occurrence. If this subsample is indeed younger than the median disk in the sample, it may not be representative of the evolution of circumstellar material in the median disk in this star-forming region.

\section{Discussion}
\label{sec:disc}

Until now, we have considered the disk mass (or luminosity) distribution of the SODA sample as a whole. There are several reasons to assume that disk masses may not be homogeneous throughout the cloud, however. Purely observationally, as demonstrated in Figure~\ref{fig:qqplot}, there is the massive tail, which may result from the superposition of a lower- and higher-mass population. Testing this hypothesis requires us to study the sample at higher resolution, and to investigate the behavior of disk masses in subsamples. By design, SODA contains a sufficiently large number of sources to make this possible. By forming subsamples of $\sim 100$ sources, we can still obtain median disk masses at similar precision to complete surveys of nearby star-forming regions. In particular, in this section we investigate the median disk mass along the length of the cloud (in Section~\ref{sec:alongcloud}) and in the different clusters of young stellar objects found in L1641 and L1647 (in Section~\ref{sec:clusters}). Finally, after investigating the presence of local variations in the median disk mass, we compare the masses of disks in SODA to those of other samples of protoplanetary disks in Orion A and B (Section~\ref{sec:SODAvsOrion}), and to nearby star-forming regions (Section~\ref{sec:otherregions}).

Two main effects may lead to spatial variations in the median mass of the protoplanetary disks on the cloud. First, if all disks evolve at similar rates and originate from the same disk mass distribution, we would expect to see mass variations tracing the median age of stars in the cloud, but this is not necessarily constant over the $\sim$ 100\,pc length of Orion A. Indeed, there is evidence for such age variations and a quite significant intrinsic age spread in Orion A~\citep{dario16}. To this end, we compare the disk masses to the ratios of Class II to embedded (Class 0 and I) and Class III YSOs, respectively, which are proxies of the mean stellar age along the line-of-sight. In this section, we use the~\citet{pillitteri13} catalog for Class III source numbers, and~\citet{tobin20} for all embedded sources (Class I and 0). Both of these surveys target only L1641. While we show results for L1647 and compare them to the embedded source numbers (derived from~\citealp{megeath12}) we are less complete to these kinds of sources there.
 
A second effect, the existence of variations in the median disk mass between different SFRs, is currently a topic of intense debate as a result of the growing number of surveys of protoplanetary disks in nearby regions. Data from the CrA and $\rho$ Oph disk populations~\citep{cazzoletti19,williams19} suggest such variations may indeed exist, but it is not clear at which spatial scale such variations occur, below the largest cloud level.

\subsection{The distribution of disk masses across the Orion A cloud}
	\label{sec:alongcloud}
\begin{figure*}
\centering
   \includegraphics[width=\textwidth]{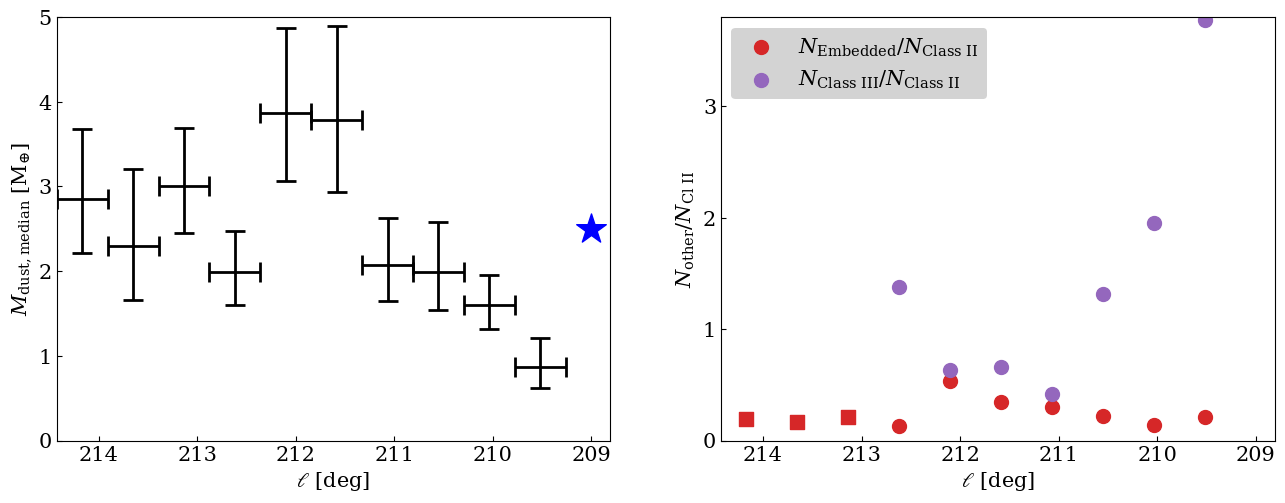}
	\caption{Trends in disk properties in SODA as a function of galactic longitude. {\it Left:} Median Class II disk masses (black points) for all longitudinal bins. The location of the Trapezium is marked with a blue star. {\it Right:} Ratios of embedded (red) and Class III YSOs (purple) to Class II sources in each bin. Round symbols show source fractions derived from~\citet{pillitteri13} and~\citet{tobin20} for Class III and embedded sources, respectively; square symbols show the embedded source fraction in L1647, derived from~\citet{megeath12}.}
     \label{fig:disksalongcloud}
\end{figure*}

	The variation of protoplanetary disk masses as a function of galactic longitude is presented Figure~\ref{fig:disksalongcloud} (left panel), where we binned the survey in seven independent longitudinal bins. Since the ``spine'' of Orion A is essentially parallel to the galactic longitude (see also Figure~\ref{fig:pointingmap}), this is a simple measure of how masses change along the cloud. Median disk masses are derived by fitting a log-normal distribution to the mass distribution of objects in each bin, with errors reflecting the errors on the fit. The widths of the distributions are highly similar, and are not shown in the Figure but presented in Table~\ref{tab:disksalongcloud}.

\begin{table*}[htbp]
\centering
	\caption{Disk properties along the spine of Orion A, by galactic longitude}
	\label{tab:disksalongcloud}
	\begin{tabular}{ c c c c c}
	\hline \hline \\[-8pt]
	$\ell$ & $N$ & $M_{\rm{dust, median} }$ & $\sigma$ & $P(A_{Ks} < 0.1)$ \\[2pt]
	$[\rm{deg}]$ &  & $[\rm{M}_{\oplus}]$ & $[\log(M_{\rm{dust}}/\rm{M}_{\oplus})]$ &  \\[2pt]
	\hline \\[-8pt]
    209.5 & 71 & $ 0.9_{ -0.2 }^{ +0.3 } $ & $ 2.0 \pm 0.3 $ & $ 0.73 \pm 0.10 $ \\[3pt]
    210.0 & 164 & $ 1.6_{ -0.3 }^{ +0.4 } $ & $ 2.1 \pm 0.2 $ & $ 0.61 \pm 0.06 $ \\[3pt]
    210.6 & 82 & $ 2.0_{ -0.5 }^{ +0.6 } $ & $ 2.0 \pm 0.2 $ & $ 0.48 \pm 0.08 $ \\[3pt]
    211.1 & 100 & $ 2.1_{ -0.4 }^{ +0.5 } $ & $ 2.0 \pm 0.2 $ & $ 0.19 \pm 0.04 $ \\[3pt]
    211.6 & 73 & $ 3.8_{ -0.9 }^{ +1.1 } $ & $ 2.1 \pm 0.2 $ & $ 0.15 \pm 0.05 $ \\[3pt]
    212.1 & 92 & $ 3.9_{ -0.8 }^{ +1.0 } $ & $ 2.1 \pm 0.2 $ & $ 0.20 \pm 0.05 $ \\[3pt]
    212.6 & 84 & $ 2.0_{ -0.4 }^{ +0.5 } $ & $ 1.8 \pm 0.2 $ & $ 0.24 \pm 0.05 $ \\[3pt]
    213.1 & 77 & $ 3.0_{ -0.6 }^{ +0.7 } $ & $ 1.6 \pm 0.2 $ & $ 0.08 \pm 0.03 $ \\[3pt]
    213.7 & 48 & $ 2.3_{ -0.6 }^{ +0.9 } $ & $ 1.9 \pm 0.3 $ & $ 0.29 \pm 0.08 $ \\[3pt]
    214.2 & 82 & $ 2.9_{ -0.6 }^{ +0.8 } $ & $ 2.0 \pm 0.2 $ & $ 0.10 \pm 0.03 $ \\[3pt]
	\hline
	\end{tabular}
\end{table*}
	
	Two results are immediately apparent from Figure~\ref{fig:disksalongcloud}. The disk masses in L1647 (corresponding to the first four bins) are statistically indistinguishable, despite spanning a significant distance of tens of parsecs when considering the three-dimensional structure of the cloud~\citep{grossschedl18}. Second, it appears that below about $\ell \sim 212^{\circ}$ degrees in galactic longitude, disk masses steadily decrease. Around $\ell = 212^{\circ}$ the disk masses seem to have a local maximum. Below $\ell = 212.6^{\circ}$, a linear decrease in median disk mass is marginally favored over a constant median disk mass, with corrected Akaike Information Criteria of AICc $= 14.2$ versus AICc $= 17.6$. Moreover, from a log-rank test the disk masses at $\ell = 209.5^{\circ}$ and $\ell = 210.0^{\circ}$ are inconsistent with those in the $\ell = 214.2^{\circ}$ and $213.1^{\circ}$ bins ($p < 0.05$). Overall, however, no large variations in the median disk mass are visible, with contrasts of a factor $\sim$few between the most and least massive bin.

	\subsubsection{Impact of age gradients}
	\label{sec:longitude}
	To test how the stellar age along the line-of-sight changes with galactic longitude, the right panel of Figure~\ref{fig:disksalongcloud} shows the fraction of Class II sources relative to embedded (Class 0, I, and flat-spectrum) and to Class III sources in each longitudinal bin. It is clear that, toward the ONC, the number of Class III sources increases sharply, while the embedded source fraction decreases. We also see a decrease in extinction $A_{Ks}$ toward these latitudes, as listed in the last column of Table~\ref{tab:disksalongcloud}. This behavior is coincident with the trend in disk masses along the cloud. Together, these observations suggest that the median age along the line of sight increases toward the ONC, as was indeed concluded by~\citet{pillitteri13}. However, if the Class III population is contributed by a population that is old enough, and distinct from the population of Class II and embedded sources, we would not necessarily expect to see any change in disk masses. Figure~\ref{fig:disksalongcloud} also shows that the embedded source fraction declines below $\ell = 212^{\circ}$. This cannot be ascribed to only a completeness effect of the {\it Spitzer} MIPS 24\,$\mu$m band, as it only saturates in the very inner part of the ONC.
	
	There is a considerable amount of literature on the physical and age structure of the young stellar populations toward Orion A which can help to shed light on the question if an older population is indeed present toward $\ell = 209.5^{\circ}$, and how it relates to the bulk of the young stellar objects in L1641. \citet{alvesbouy12} and~\citet{bouy14} suggested a large population of stars associated with NGC 1980 but kinematically similar to L1641 is present in this area, with an age of 5 -- 10\,Myr. However, the existence of such a population has been challenged since then:~\citet{dario16} suggest that while older stars are present and have lower extinctions, they are not completely separate from the cloud. On the other hand, an older foreground is not retrieved at all by~\citet{fang17} and~\citet{kounkel17}, who conclude that the foreground population is indistinguishable both in terms of kinematics and age. A population of stars belonging kinematically to Orion D may be present in this area per~\citet{kounkel18}. A similar group of stars is identified by~\citet{zari19} and in~\citet{chen20}; these suggested populations have ages in the $\sim 10$\,Myr range. Combining these results with a three-dimension extinction map,~\citet{rezaei20} conclude that the evidence favors an older foreground population along these lines of sight.
	
	Crucially, and regardless of precise three-dimensional structure, the age ranges quoted for the older part of the stellar population in this area of Orion A suggest that we can expect  $\sim 10\%$ of these stars to have a near-infrared excess that would lead to these stars being included in this survey~\citep{hernandez08,luhman22}. At the same time, disks at this age are faint at millimeter continuum wavelengths~\citep{barenfeld16,ansdell20}. Depending on the size of the older population, then, we can explain the behavior observed in Figure~\ref{fig:disksalongcloud}: along the line-of-sight toward $\ell = 209^{\circ}$, the median age and the width of the age distribution are larger than at higher latitudes, so the fraction of embedded sources to Class II objects decreases, while the Class III to Class II ratio climbs.
	
	To explore the hypothesis that a mixed population along the line of sight leads to a lower median disk mass and higher Class III to Class II ratio more quantitatively, we artificially construct a sample of disks by randomly sampling (with replacement) disks from both the $\ell > 213.1^{\circ}$ area in Orion, and from the Upper Sco catalog published by \citet{barenfeld16}. This approach thus requires the assumption that disk evolution proceeds similarly in these star-forming regions. For a ratio $f = 0.68 - 0.84$ between these populations, the median disk mass of the mixed sample is indistinguishable from $\ell = 209^{\circ}$ disks, to within $1\sigma$. This approach is shown in Figure~\ref{fig:popmix}. Using a Class III to Class II ratio for Upper Sco of 4.2~\citep{luhman22} and $\sim 1$ for the southern L1641 sample leads to an $N_{III}/N_{II}$ ratio of $\sim 2$ for the mixed sample, similar if slightly lower than that observed. However, it is important that the fraction of diskless young stars is determined in different ways in both regions, adding additional uncertainty to this estimate. Ultimately, the limited statistical significance of the median disk mass variations in L1641, the assumptions on the disk mass distribution for 5--10\,Myr old disks, and in particular the uncertainties in the (relative) stellar ages of the stars with protoplanetary disks in L1641 prevent us from drawing stronger conclusions.
	
	Alternatively, the disk masses in the lowest galactic longitudes might be affected by external photoevaporation. This process is expected to be an important cause for disk mass loss in the immediate vicinity of the Trapezium~\citep{mann14,eisner18}. However, it seems to be less important even at only $\sim 2$\,pc distances in the OMC-2 region~\citep{vanterwisga19b}. The SODA sample is separated by at least 4\,pc from the Trapezium. However, it is possible that the proximity of $\iota$ Ori (an O9III star) to L1641N does contribute to lowering the masses of the disks there.

	\subsubsection{Disk masses in YSO clusters}
	\label{sec:clusters}
	As is clear from the distribution of sources in Figure~\ref{fig:pointingmap}, Class II YSOs are not spread out uniformly across the Orion A cloud. Even in our sample, which focuses explicitly on the disks located away from the ONC, several regions with higher densities are readily apparent by eye, and such clusters have been identified in this sample systematically by~\citet{megeath16}.
	
	Considering the properties of disks from the perspective of these associations gives us valuable information. Even disregarding the possible foreground contamination in the northern part of our survey area, stars currently in overdensities should be more similar to each other than stars not in overdensities -- a result that was previously demonstrated in~\citet{megeath16}. The field population can, after all, have originated in earlier, now dissolved, overdensities, and should thus at least have a larger age dispersion~\citep{kainulainen17}. Moreover, clusters are less likely to have significant structure along the line-of-sight.
	
	We restrict ourselves to finding clusters in two-dimensional (RA, Dec) space, as a first-order approach that has been successfully used for this dataset before. Even with {\it Gaia} EDR3 the fraction of sources with accurate astrometry is fairly low: 284/873, using version 1.0 of the astrometric fidelity classifier presented in~\citet{rybizki21}, which means this is the most practical approach that ensures sufficiently large sample sizes. To search for clusters, we used the OPTICS algorithm~\citep{optics99}. The data distances were scaled, and we required a minimum of five samples in the neighborhood for core points to be considered, as well as a minimum reachability steepness of $\xi = 0.05$. We additionally set a minimum cluster size of 50 members of Class II or earlier. This choice of parameters allows us to minimize the number of interlopers, while retrieving similar clusters to those reported originally in~\citet{megeath16}, and ensures accurate estimates of the median disk masses of the cluster members. While we use the term cluster, we do not expect any of these to be gravitationally bound to any significant degree. The stellar surface densities of these clusters are too low for that, with between 10 -- 30\,YSOs\,pc$^{-2}$~\citep{megeath16}. 
	
\begin{figure*}
\sidecaption
  \includegraphics[width=12cm]{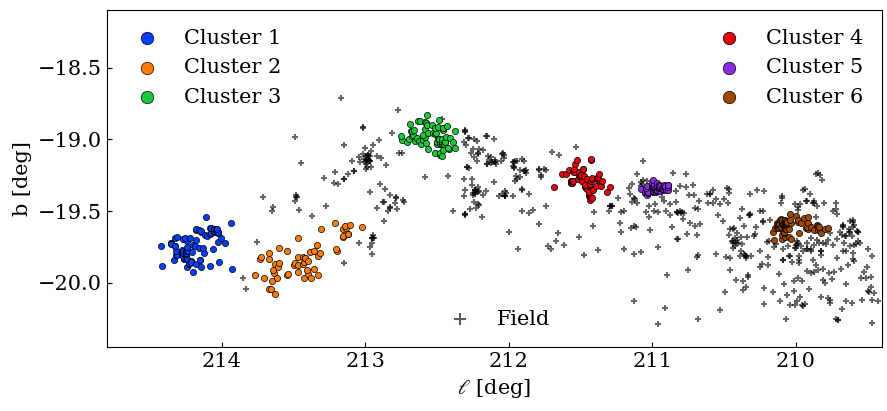}
     \caption{Clusters identified by OPTICS. Cluster members are indicated by colored circles, while field stars are shown as gray plus signs.}
     \label{fig:clusteronsky}
\end{figure*}

	The resulting 6 clusters and the unclustered field population are shown in Figure~\ref{fig:clusteronsky}. It is clear that sufficiently large overdensities of YSOs are present throughout the cloud. However, toward L1641N, the ratio of cluster members to field stars begins to drop rapidly. We probe a range of surface (over)densities, with Cluster 2 being the least dense. Compared to the clusters found in~\citet{megeath16}, who used a different cluster-finding approach, we see that they do not retrieve cluster 5, while cluster 2 is broken up into two smaller clusters, one of which is (significantly) smaller. Since~\citet{megeath16} do not require a minimum cluster size, they also find several other small clusters and groups throughout.

\begin{figure*}
\centering
   \includegraphics[width=\textwidth]{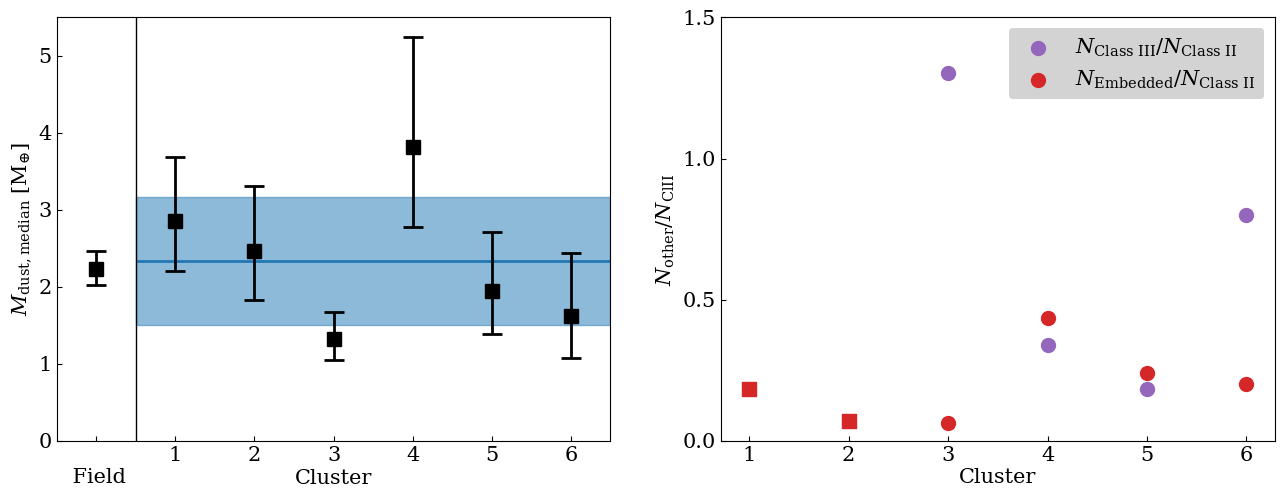}
  \caption{Median disk dust masses in clusters and the field population. {\it Left:} median masses for the individual clusters and field population (black) and the median and standard deviation for the clusters together. {\it Right:} ratio of the numbers of embedded (red) and Class III YSOs (purple) to Class II sources in each cluster. Round symbols show source fractions derived from~\citet{pillitteri13} and~\citet{tobin20} for Class III and embedded sources, respectively; square symbols show the embedded source fraction in L1647, derived from~\citet{megeath12}.}
  \label{fig:clustermass}
\end{figure*}

\begin{table*}[htbp]
	\centering
	\caption{Disk properties in YSO clusters identified in this sample}
	\label{tab:clusters}
		\begin{tabular}{cccccc}
		\hline \hline \\[-8pt]
		Cluster & $\ell$ & $N$ & $M_{\rm{dust, median} }$ & $\sigma$ & $P(A_{Ks} < 0.1)$ \\[2pt]
		 & $[\rm{deg}]$ &  & $[\rm{M}_{\oplus}]$ & $[\log(M_{\rm{dust}}/\rm{M}_{\oplus})]$ & \Tstrut\Bstrut\\ \hline
        1 & 214.2 & 82 & $ 2.9_{ -0.6 }^{ +0.8 } $ & $ 2.0 \pm 0.2 $ & $ 0.10 \pm 0.03 $ \Tstrut\Bstrut\\
        2 & 213.4 & 56 & $ 2.5_{ -0.6 }^{ +0.8 } $ & $ 1.9 \pm 0.3 $ & $ 0.18 \pm 0.06 $ \Tstrut\Bstrut\\
        3 & 212.5 & 63 & $ 1.3_{ -0.3 }^{ +0.3 } $ & $ 1.5 \pm 0.2 $ & $ 0.30 \pm 0.07 $ \Tstrut\Bstrut\\
        4 & 211.5 & 53 & $ 3.8_{ -1.0 }^{ +1.4 } $ & $ 2.2 \pm 0.3 $ & $ 0.06 \pm 0.03 $ \Tstrut\Bstrut\\
        5 & 211.0 & 54 & $ 1.9_{ -0.6 }^{ +0.8 } $ & $ 2.1 \pm 0.3 $ & $ 0.06 \pm 0.03 $ \Tstrut\Bstrut\\
        6 & 210.1 & 50 & $ 1.6_{ -0.5 }^{ +0.8 } $ & $ 2.3 \pm 0.4 $ & $ 0.46 \pm 0.10 $ \Tstrut\Bstrut\\ \hline
        Field & 211.1 & 515 & $ 2.2_{ -0.2 }^{ +0.2 } $ & $ 2.0 \pm 0.1 $ & $ 0.43 \pm 0.03 $ \Tstrut\Bstrut\\
		\hline
		\end{tabular}
\end{table*}

	In Figure~\ref{fig:clustermass}, the median disk masses as inferred from a log-normal fit to the disk mass distribution of each cluster's members are presented, as well as the average disk mass for all field members. Additionally, Table~\ref{tab:clusters} shows the number of members, median masses, and widths of the log-normal distributions for the clusters and field. The latter is not significantly different from the global average, which is not surprising, since field-disks (508) outnumber cluster-disks by a significant amount, with a median of $59.5$ members per cluster.
	
	Figure~\ref{fig:clustermass} shows that the differences between disk masses in clusters are not particularly large, like those seen when longitudinally binning the disks along the cloud. However, some variation does exist. Clusters 1 and 2, both located in L1647, have similar median disk masses, and are each statistically indistinguishable from the field. Together, these clusters contain the majority of the Class II sources in L1647 (138 of 165).

	In L1641, clusters 3, 5, and 6 have median disk masses that are again similar. For all three of these clusters, however, the median disk mass is lower than the field, although this is significant only for cluster 3. Since clusters 5 and 6 are coincident with the part of L1641 where, above, we concluded overall ages may be higher, it is perhaps not surprising that their median disk masses are also lower, especially since we projected the three-dimensional structure of the cloud into two dimensions. It is consistent, in other words, with the cluster approach partially compensating for a (more dispersed, older) foreground population.
	
	Comparing the clusters to the embedded and Class III object populations provides additional insight into what causes the differences between the densest groups of Class II YSOs. Again, we see that the ratio of Class III to Class II sources varies significantly. Clusters 3 and 6 coincide with relatively high numbers of Class III objects, which may provide an explanation for their lower disk masses relative to the field. The differences in $N_{\textrm{embedded}} / N_{\textrm{Class II}}$ and $N_{\textrm{Class III}} / N_{\textrm{Class II}}$ between clusters 4 and 5 are small. Cluster 4 contains relatively more young, embedded objects. This may be seen as evidence that it is the youngest of the clusters we identified here, and could explain at least in part the relatively high median disk mass of this cluster.
	
	We must stress that the differences between these clusters, which are located far from each other on the sky, are small and not highly significant relative to each other and to the field, even in the most extreme cases. In the surveyed part of Orion A, where we avoid the presence of O-stars, it appears to be the case that disk evolution as traced by millimeter-continuum emission proceeds similarly throughout the cloud when studied with sample sizes comparable to those found for nearby SFRs, and especially when taking into account the fractions of embedded- and Class III sources as a proxy for age. If variations in the initial disk mass (or rate of evolution) in the part of Orion A surveyed here do exist, they are not significant with the present analysis, although we cannot exclude their existence completely.
	
	These observations are consistent with the expectations from theoretical work: in environments like the low-mass, low-density clusters we are probing here, the disk evolution models run by~\citet{concharamirez19} predict no impact from photoevaporation on the disk population. This effect is made even stronger by the lack of O-type stars in this part of Orion A~\citep{hsu12,hsu13}. In the following, we compare the SODA sample to other nearby star-forming regions, in order to see if the observed similarity between different groups of young disk-bearing stars holds at scales larger than GMCs.

	\subsection{Disk masses: SODA and other star-forming regions}
	\label{sec:otherregions}
	As the previous discussion has shown, the stellar population of the parts of Orion A probed by our survey -- from L1647 to NGC1980 -- does not have a single age distribution. Surface overdensities reveal some older or younger populations, and there is a trend toward higher ages with decreasing galactic longitude in the cloud as a whole. In each case, this is reflected in the disk mass distribution: older populations have lower median disk masses. In contrast, in those regions where stellar ages appear to be most homogeneous, and in particular in the southern half of L1641S, we do not see significant changes in disk masses. This implies that disk masses in a region like L1647 only depend on the age, and perhaps the global cloud properties on scales of at least several tens of parsecs.
	
	The logical extension of this idea is to compare the masses of disks with similar ages between different SFRs, which we do here, to see if disk properties are the same at similar ages even in quite different SFRs. This requires the careful selection of a subsample of Class II objects from SODA. Because of the complicated age structures of the stellar population in L1641N, we focus primarily on L1641S and L1647. However, the cluster analysis in section~\ref{sec:clusters} shows that even in L1641S, outliers in age (and disk mass) are present. We therefore define a sample we call L1641S (reduced). This consists of all YSOs with $\ell > 212.8$, and all YSOs with $211.8 \leq \ell < 212.3$. This means that both cluster 3 and cluster 4 are avoided, and presents a sample with the most uniform age- and disk mass distributions, based on information from the number of Class III and embedded sources in the field. For this sample, we adopt an age of $1.5 \pm 0.7$\,Myr, based on~\citep{dario16}. At the same time, we define L1641N to consist of all disks in our sample with $\ell < 211.2$. This sample has a higher age, and -- potentially -- a wider age spread, due to the uncertainties in the stellar population structure here.

\begin{figure}
  \resizebox{\hsize}{!}{\includegraphics{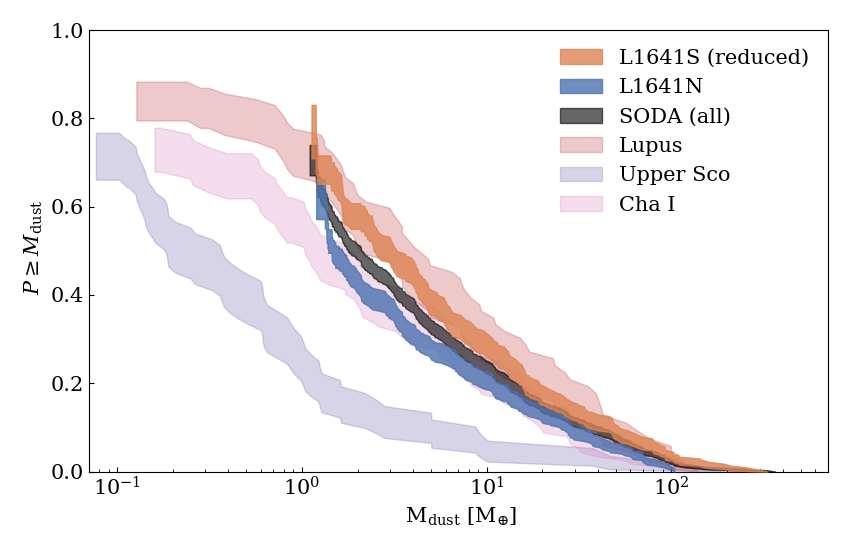}}
  \caption{Comparison of the SODA results to three nearby SFRs. Kaplan-Meier estimators for the cumulative disk mass distribution in L1641S (reduced) (orange), L1641N (blue), the full SODA sample (gray), and three literature regions: Lupus (tan), Cha I (pink), and Upper Sco (purple)\citep{ansdell16, pascucci16,barenfeld16}.}
  \label{fig:kmfNS}
\end{figure}

	\subsubsection{Nearby star-forming regions}
	\label{sec:SODAvsother}
	In Figure~\ref{fig:kmfNS}, the Kaplan-Meier estimators for the disk mass distribution function in L1641S (reduced), L1641N, and the full SODA sample are shown, as well as the disk mass distributions for three nearby star-forming regions with different ages: Lupus (1--3\,Myr, \citealt{sfr_book}), Cha I (2--3\,Myr, \citealt{luhman08}), and Upper Sco (5--10\,Myr, \citealt{slesnick08}). Remarkably, the mass distributions for disks in L1641S (reduced) and Lupus are indistinguishable statistically ($p < 0.05$). On the other hand, the L1641N disks more closely resemble the older Cha I population, albeit marginally ($p = 0.04$) in a log-rank test, although we caution that the comparison between these populations may not be entirely proper given the uncertainties in the stellar age spread of the L1641N population, which is likely to be larger than that in Chamaeleon I. 
	
	The main conclusion that can be drawn from this comparison is that the disks in SODA, when taking into account the age gradient along the spine of Orion A, have similar masses at similar ages as the protoplanetary disks in entirely unrelated, nearby low-mass SFRs. This in turn implies that their initial conditions and subsequent evolution must have been similar. If nothing else, this suggests that the results obtained from high-resolution observations of the disk continuum in these regions can be generalized to those regions of Orion A where no photoionizing O-type stars are found.

\begin{figure*}
  \resizebox{\hsize}{!}{\includegraphics{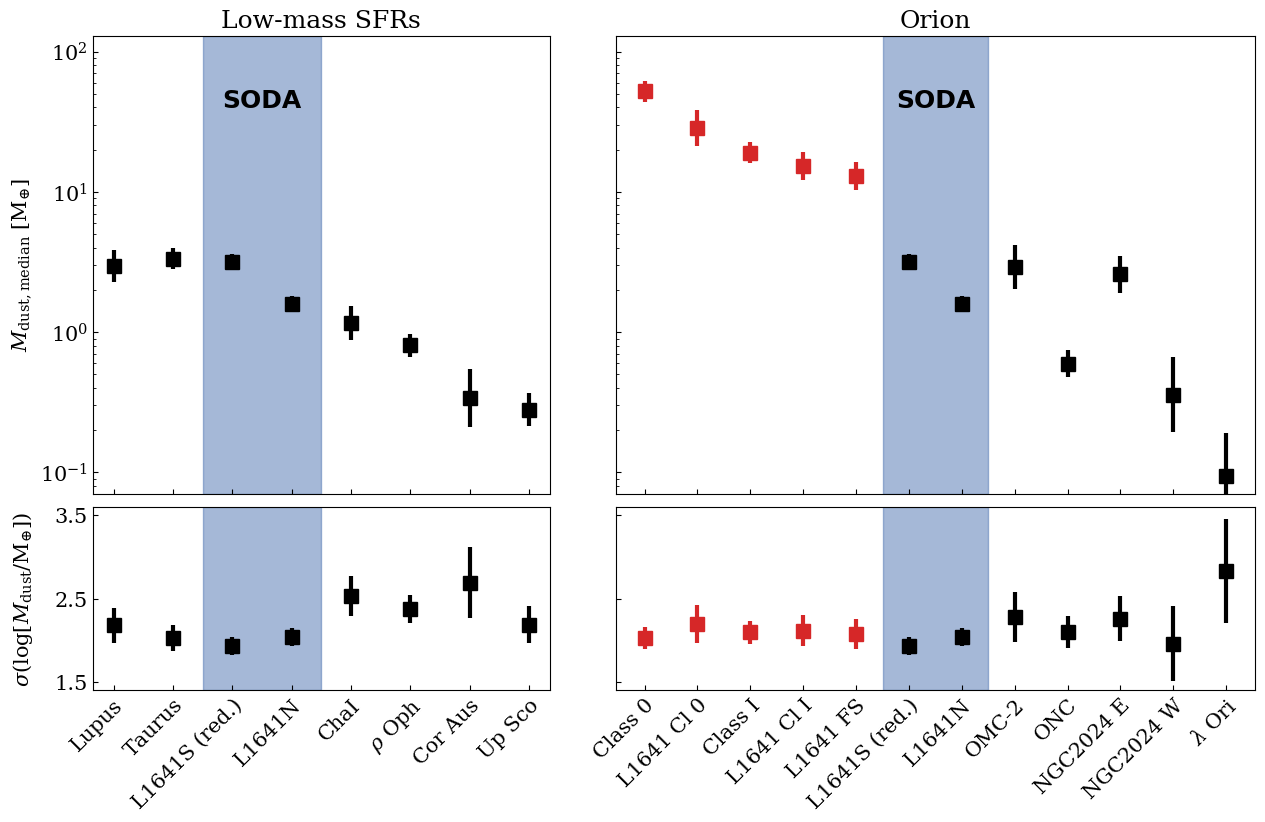}}
  \caption{Properties of the disk dust mass distributions for disks in nearby star-forming regions and Orion. Median masses (top row) in $M_{\oplus}$ and widths of the log-normal distributions (bottom row), in $\log(M_{\rm{dust}}/M_{\oplus})$. Left, nearby star-forming regions, sorted by median mass: Lupus \citep{ansdell16}, Taurus \citep{andrews13}, L1641S (reduced), L1641N, Cha I \citep{pascucci16}, $\rho$ Oph \citep{williams19}, CrA \citep{cazzoletti19}, and Upper Sco \citep{barenfeld16}. Right, different populations in Orion A and B: Class 0 and Class I \citep{tobin20} in the full sample and L1641 (in red), L1641S (reduced) and L1641N, OMC-2 \citep{vanterwisga19b}, the ONC \citep{eisner18}, NGC2024 E and W \citep{vanterwisga20}, and $\lambda$ Ori~\citep{ansdell20}.}
  \label{fig:allclouds}
\end{figure*}

\begin{table}[htbp]
\caption{Median disk masses and log-normal widths for nearby star-forming regions}
\label{tab:allclouds}
\begin{tabular}{lccc}
\hline \hline
SFR & N & $M_{\rm{dust,median}}$ & $\sigma$ \\
    &   &  $[M_{\oplus}]$     & $[\log(M_{\rm{dust}}/ \rm{M}_{\oplus})]$ \Tstrut\Bstrut \\ \hline 

SODA (all)  & 873 & $2.2^{+0.2}_{-0.2}$ & 2.0$\pm$0.1 \Tstrut\Bstrut \\
L1641 (all) & 708 & $2.1^{+0.2}_{-0.2}$ & 2.0$\pm$0.1 \Tstrut\Bstrut \\
L1647 (all) & 165 & $2.6^{+0.4}_{-0.5}$ & 1.9$\pm$0.2 \Tstrut\Bstrut \\ \hline

L1641S (red.) & 301 & $3.2^{+0.4}_{-0.4}$ & 1.9$\pm$0.1 \Tstrut\Bstrut \\
L1641N & 407 & $1.6^{+0.2}_{-0.2}$ & 2.0$\pm$0.1 \Tstrut\Bstrut \\ \hline 

Lupus & 69 & $3.0^{+0.7}_{-0.9}$ & 2.2$\pm$0.2 \Tstrut\Bstrut \\
Taurus & 178 & $3.3^{+0.5}_{-0.6}$ & 2.0$\pm$0.2 \Tstrut\Bstrut \\
ChaI & 93 & $1.2^{+0.3}_{-0.4}$ & 2.5$\pm$0.2 \Tstrut\Bstrut \\
$\rho$ Oph & 172 & $0.8^{+0.1}_{-0.2}$ & 2.4$\pm$0.2 \Tstrut\Bstrut \\
Cor Aus & 43 & $0.3^{+0.1}_{-0.2}$ & 2.7$\pm$0.4 \Tstrut\Bstrut \\
Up Sco & 75 & $0.3^{+0.1}_{-0.1}$ & 2.2$\pm$0.2 \Tstrut\Bstrut \\ \hline

Orion Class 0 & 133 & $52^{+8.4}_{-10}$ & 2.0$\pm$0.1 \Tstrut\Bstrut \\
L1641 Class 0 & 56 & $28.7^{+7.3}_{-9.9}$ & 2.2$\pm$0.2 \Tstrut\Bstrut \\
Orion Class I & 150 & $19.0^{+3.0}_{-3.6}$ & 2.1$\pm$0.1 \Tstrut\Bstrut \\
L1641 Class I & 89 & $15.2^{+3.1}_{-3.9}$ & 2.1$\pm$0.2 \Tstrut\Bstrut \\
L1641 Flat Spectrum & 83 & $13.0^{+2.7}_{-3.4}$ & 2.1$\pm$0.2 \Tstrut\Bstrut \\
OMC-2 & 132 & $2.9^{+0.9}_{-1.3}$ & 2.3$\pm$0.3 \Tstrut\Bstrut \\
ONC & 226 & $0.6^{+0.1}_{-0.1}$ & 2.1$\pm$0.2 \Tstrut\Bstrut \\
NGC2024 E & 97 & $2.6^{+0.7}_{-0.9}$ & 2.3$\pm$0.3 \Tstrut\Bstrut \\
NGC2024 W & 82 & $0.36^{+0.2}_{-0.3}$ & 2.0$\pm$0.5 \Tstrut\Bstrut\\
$\lambda$ Ori & 44 & $0.09^{+0.05}_{-0.10}$ & 2.8$\pm$0.6 \Tstrut\Bstrut\\ \hline
\end{tabular}
\end{table}

	\subsubsection{Star-forming regions in Orion A and B}
	\label{sec:SODAvsOrion}
	Figure~\ref{fig:allclouds} extends this comparison between the SODA sample and Class II disk masses in other nearby star-forming regions, as well as the Class 0 and Class I disk mass distributions in Orion A and B presented by~\citep{tobin20}. The same quantities are also shown in Table~\ref{tab:allclouds}. Log-normal distributions were fit to all the observed disk populations, and their medians and standard deviations (in $\log(M_{\rm{dust} / M_{\oplus}})$ are shown. This time, we plot masses on a logarithmic scale, unlike the linear scales in Figures~\ref{fig:disksalongcloud} and~\ref{fig:clustermass}, to accommodate the wide range. The left panel focuses on the comparison of the SODA sample to nearby low-mass star-forming regions, where no external photo-evaporation is expected to play an important role. From this panel, we exclude IC348~\citep{ruizrodriguez18}, which is dominated by low-mass stars, and thus has lower-than-expected disk masses if not corrected for its bottom-heavy IMF. The right panel of this figure shows the disk masses of a number of previously-studied regions in Orion A and B. Here, only $\sigma$ Ori~\citep{ansdell17}, which is not well-described by a (single) log-normal distribution due to significant external photoevaporation of disks in the cluster center, is excluded.
	
	Several important conclusions can be drawn from this figure. First, it is clear that the median disk mass of the L1641S (reduced) sample is indistinguishable from that in OMC-2, almost on the opposite side of Orion A from the Trapezium. It also closely resembles not only Lupus~\citep{ansdell16}, but also Taurus~\citep{andrews13}, both of which have a similar age~\citep{sfr_book,luhman04} and similar median disk masses.
	
	Comparison between the median disk masses in SODA and other parts of Orion is complicated by the wide range of environmental conditions in these clouds. While the OMC-2 is similar to the sample studied here, and not affected by O-type stars, disks in the ONC are subject to significant amounts of external photoevaporation due to the proximity of the Trapezium stars. The same is true for the NGC2024 W disks, and recently~\citet{haworth21} also showed that proplyds are present in NGC2024 E -- the youngest sample of disks in this figure. $\lambda$ Ori, meanwhile, is known to be an older region overall ($\sim 5$\,Myr, \citealt{dolan01,mathieu08}), and indeed less massive than any other population plotted here, regardless of the presence of external photoevaporation.
	
	\subsubsection{Mass evolution from Class I to Class II in Orion}
	By combining SODA with the results from the VANDAM survey of Orion by~\citet{tobin20}, we have the most precise picture yet of how protoplanetary disk masses evolve from the earliest embedded phases to Class II (Fig.~\ref{fig:allclouds}). As these authors also showed, the most embedded Class 0 objects may be lower in mass in L1641 (their sample does not cover L1647). On the other hand, Class I disks do not show significant differences between L1641 and the full sample of such objects in Orion A and B. This suggests that YSOs in L1641 are representative of (nonphotoevaporating) environments in Orion also at younger ages. Thus, we can use these sample to study the mass evolution of disks in a general sense.
	
	From Class I to Class II, the disk mass is reduced by a factor of $13 \pm 3$, based on these observations. If we assume the evolution from Class I to Class II takes about 1\,Myr and continue to use a gas-to-dust mass ratio of 100, that implies an average disk mass loss rate of $\dot{M} \sim 10^{-8}\,M_{\odot}$\,yr$^{-1}$ during this period. This number represents the net outcome of a number of different processes: the growth of dust to pebbles and potentially larger bodies which are not detected at these wavelengths, the accretion of material onto the star and its removal through outflows, and the accretion of material from the envelope toward the disk.
	
	\citet{furlan16} find values of $10^{-6} - 5\times10^{-7}\,M_{\odot}$\,yr$^{-1}$ for the infall rate of new material from the envelope onto the disk and star in the Class I and flat-spectrum sources from SED modeling. However, as these authors suggest, these models may not realistically reproduce the envelope structure, and the stellar masses were not known in these models but are a key parameter for estimates of the infall rate. Thus, they may significantly overestimate the amount of material accreted onto the disk. Indeed, direct observations of inverse P-Cygni profiles associated with infall from envelope species like H$_2$O and HCO$^+$ are remarkably rare in Class I sources~\citep{kristensen12,mottram17,vandishoeck21}. There is evidence for significant grain growth even in the most embedded phases of protostellar evolution~\citep{harsono18}, making it difficult to quantify the efficiency of this process from Class I to II. \citet{tychoniec20} suggested that planet formation occurring as early as Class 0 may help to explain the observed masses of exoplanetary systems; this would also lead to lower millimeter-dust emission at a rate of approximately $10^{-9}\,M_{\odot}$\,yr$^{-1}$. Interestingly, the rate of mass loss from the disk from Class I to II is of the same order of magnitude as the measured values of the stellar accretion rate for Class II objects~\citep{alcala17} (which in turn are approximately a factor of 10 larger than the mass loss rates via jets~\citep{cabrit07,frank14}). Combining this observation with the statistically indistinguishable masses of Class I and flat-spectrum disks, we might infer that the observed disk mass at these ages is in (near) steady-state, as also suggested by~\citet{tsukamoto17}. In that case, our closed-box estimate of mass loss describes the evolution from the end of significant envelope accretion, as early as Class I, to the $\sim$ Myr-old populations in SODA. However, more detailed studies -- in particular observations at higher resolutions of substructures in the youngest disks, and of the interaction between envelope material and the disk -- are clearly needed to fully disentangle these competing processes in the (partially) embedded phases and beyond.
	
	\subsubsection{Existence of a disk mass - age relation in isolation}
	The similarity in median disk masses across several distinct populations of Class II YSOs is intriguing. However, it is not without exceptions. Leaving aside IC 342, the two regions that have been suggested in the literature to have low disk masses for their age are CrA and $\rho$ Oph~\citep{williams19,cazzoletti19}. For CrA, multiple authors have found young ages (in the 1--2\,Myr range, ~\citealt{sicilia-aguilar11,cazzoletti19}). However, evidence from {\it Gaia} now suggests ages of 5--6\,Myr for at least a significant fraction of its population~\citep{galli20}. This suggests that this region's disks should more closely resemble those in Upper Sco in terms of mass, but it is not clear if this older population is large enough to fully explain this region's disk masses. 
	
	In $\rho$ Oph, there is similar evidence from {\it Gaia} that its YSO population is contaminated to at least some extent by an older, kinematically distinct population, with an age of 5--10\,Myr~\citep{grasser21}. However, again, it is not clear that this population is large enough to fully explain the low observed disk masses. From comparison with L1641N, we can see that it is at least plausible that such a combined population would have median disk masses in the range observed. If this is not the case, however, it suggests that some large-scale property of the $\rho$ Oph SFR must be responsible for the low masses of disks there.
	
	Summarizing, the majority of evidence indicates that disk masses in a given SFR are the same at similar ages (in the absence of external photo-evaporation). Outliers like $\rho$ Oph and CrA seem to be rare, both in absolute numbers of disks and in terms of different nearby SFRs. If this is the case, it should be possible to use the nearby SFRs as a representative testing ground for theories of disk formation and evolution.
	
	At the same time, this work places strong constraints on the area over which the properties of disks can vary. Compared to previous surveys, the results presented here show for the first time that even within the tens of parsecs spanned by Orion A, no large variations in median disk masses occur, especially after considering the impact of the overall variations in the ages of the sample. This means that if the disks in the outlier regions such as $\rho$ Oph, CrA, and perhaps L1641N are indeed intrinsically less massive, we must look for the cause at the level of giant molecular clouds. Large-scale cosmic ray ionization trends have been proposed~\citep{kuffmeier20}, while cloud-scale disk evolution models seem to suggest that turbulence is more important than magnetic fields on the largest scales~\citep[e.g.,][]{kuffmeier17, bate18}. 

\section{Conclusions}
We have conducted an unbiased survey at 225 GHz of 873 Class II disks identified by {\it Spitzer} in the southern parts of Orion A, the largest such survey as of yet, by a factor of a few compared to previous work. These observations make it possible to characterize the disk mass distribution of the entire southern part of Orion A as a whole, and enable us to determine the disk mass distribution for clusters of YSOs in the cloud. SODA has unlocked a large, well-characterized sample of protoplanetary disks, which closely resembles smaller nearby star-forming regions. The power of this sample size allows us to distinguish the subtle effects that govern the evolution of protoplanetary disk material at the population level.

Using these data, we have studied the evolution of disk masses, both within the cloud, and in nearby low-mass star-forming regions. We also provide, for the first time, a complete catalog of the most luminous protoplanetary disks at submillimeter wavelengths in L1641 and L1647. We identified an object that may be actively accreting new circumstellar material in sufficient quantities for the accretion stream to be detectable in the 225\,GHz continuum, and determine that such objects are rare ($\sim 0.1\%$ of all sources); alternatively, it may be a more embedded object seen (nearly) face-on.

Our main conclusions are as follows:

	\begin{enumerate}
		\item Of the 873 Class II disks that make up the SODA sample, we detect 502, or 58$\%$ of the full sample. This detection rate lies in between those for Chamaeleon I and Lupus at the same sensitivity.
		\item We identify a sample of 20 disks with masses larger than $100\,\textrm{M}_{\oplus}$, more than doubling the number of such disks in L1641 alone. This is the largest unbiased sample of massive protoplanetary disks in a single cloud to date, and ideally suited to high-resolution follow-up observations.
		\item By fitting a log-normal distribution to the observations, we find median disk dust masses of $2.2^{+0.2}_{-0.2}\,M_{\oplus}$ (full sample) $2.1^{+0.2}_{-0.2}\,M_{\oplus}$ (L1641) and $2.6^{+0.4}_{-0.5}\,M_{\oplus}$ (L1647) respectively.
		\item toward the northern part of L1641, the disk masses apparently decrease, even as the fraction of Class III sources increases and embedded sources become rarer. This is consistent with earlier results showing an older foreground population toward this region.
		\item We identify six clusters of YSOs with a sufficiently large number of cluster members that their disk properties can be compared statistically. The average disk masses of the YSO clusters we identify in Orion A show differences that are driven primarily by their age.
		\item In L1641 S, where we can define a large sample of YSOs with similar ages, we find that disk masses are indistinguishable from those in nearby low-mass star-forming regions in Lupus and Taurus, which have similar ages.
		\item  Together, these results suggest that disk masses are the same at similar ages (in the absence of external photoevaporation) for many, if not all, nearby SFRs, and that they are similar -- and evolve similarly -- at scales of $\sim 100$\,pc in Orion.
	\end{enumerate}

The remarkably homogeneous properties of disk samples of the same age are a surprising finding, already hinted at by previous surveys of nearby star-forming regions, which found that disk mass distributions in the Lupus and Taurus star-forming regions are statistically indistinguishable. Now, however, we can show this applies to a larger number of YSOs and YSO clusters, forming in well-separated parts of the same giant cloud. For the first time, the unprecedented size of the SODA disk sample allows us to zoom in on the effects of age gradients and clustering in a single star-forming region. By combining fast ALMA mapping techniques and parallel data processing developed for this study, we open a new window on the evolution of protoplanetary disks.

\begin{acknowledgements}
We thank Sierra Grant for insightful discussions on the disk populations in Orion. This paper makes use of the following ALMA data: ADS/JAO.ALMA\#2019.1.01813.S. ALMA is a partnership of ESO (representing its member states), NSF (USA) and NINS (Japan), together with NRC (Canada), MOST and ASIAA (Taiwan), and KASI (Republic of Korea), in cooperation with the Republic of Chile. The Joint ALMA Observatory is operated by ESO, AUI/NRAO and NAOJ.
This project has received funding from the European Research Council (ERC) under the European Union’s Horizon 2020 research and innovation programme (Grant agreement No. 851435).
This work made use of the Dutch national e-infrastructure with the support of the SURF Cooperative using using grant no. EINF-259.
The authors acknowledge assistance from Allegro, the European ALMA Regional Center node in the Netherlands.
\end{acknowledgements}

\bibliographystyle{aa}
\bibliography{BigBib}

\begin{appendix}
\section{Potentially resolved sources}

\begin{figure*}[b]
\centering
   \includegraphics[width=16cm]{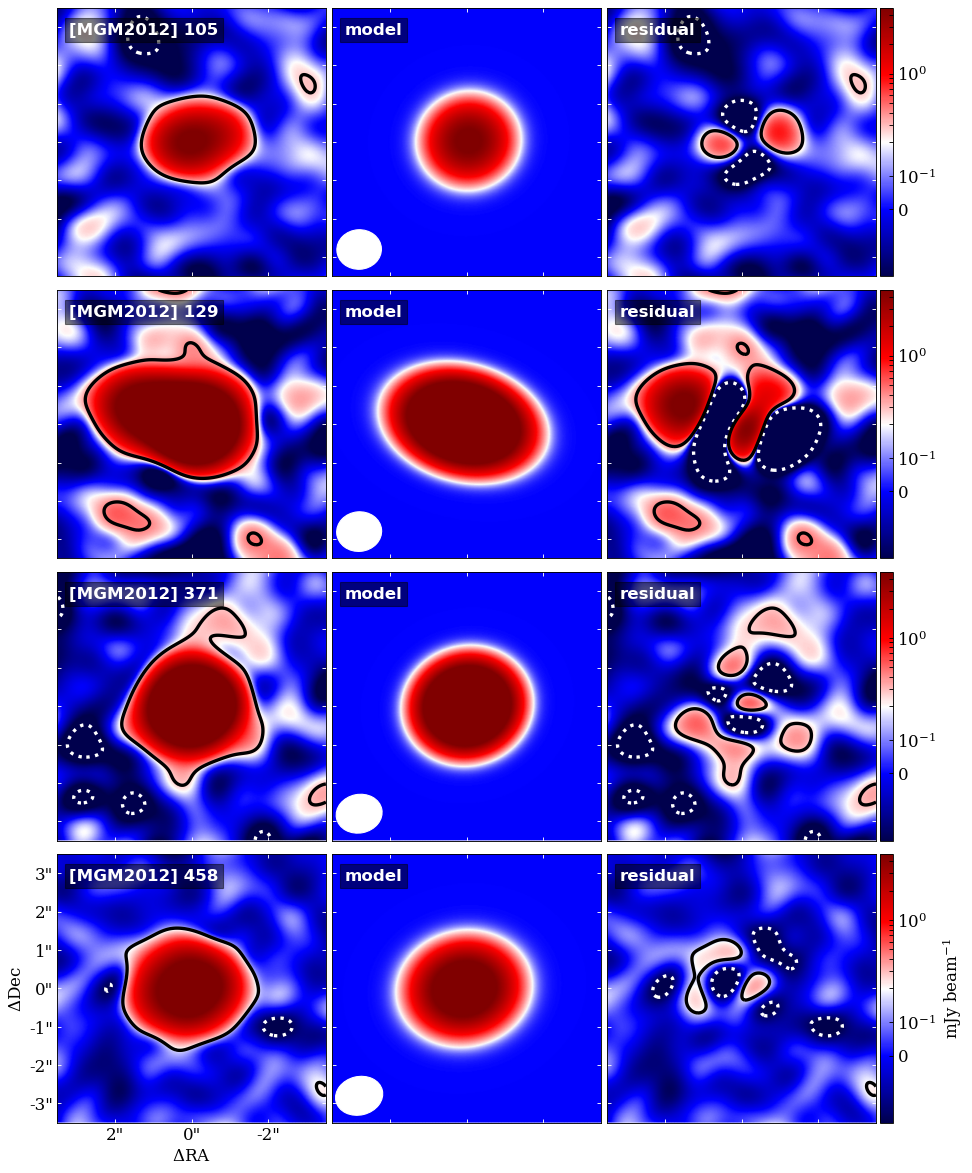}
     \caption{Sources with evidence of resolved emission from double Gaussian fitting. Data, model, and residuals are shown per source from left to right. Contours indicate the $+4\sigma$ (solid black) and $-4\sigma$ (dashed white) contours. The beam is shown in white in the middle image.}
     \label{fig:resolved_0}
\end{figure*}

\begin{figure*}
\centering
   \includegraphics[width=16cm]{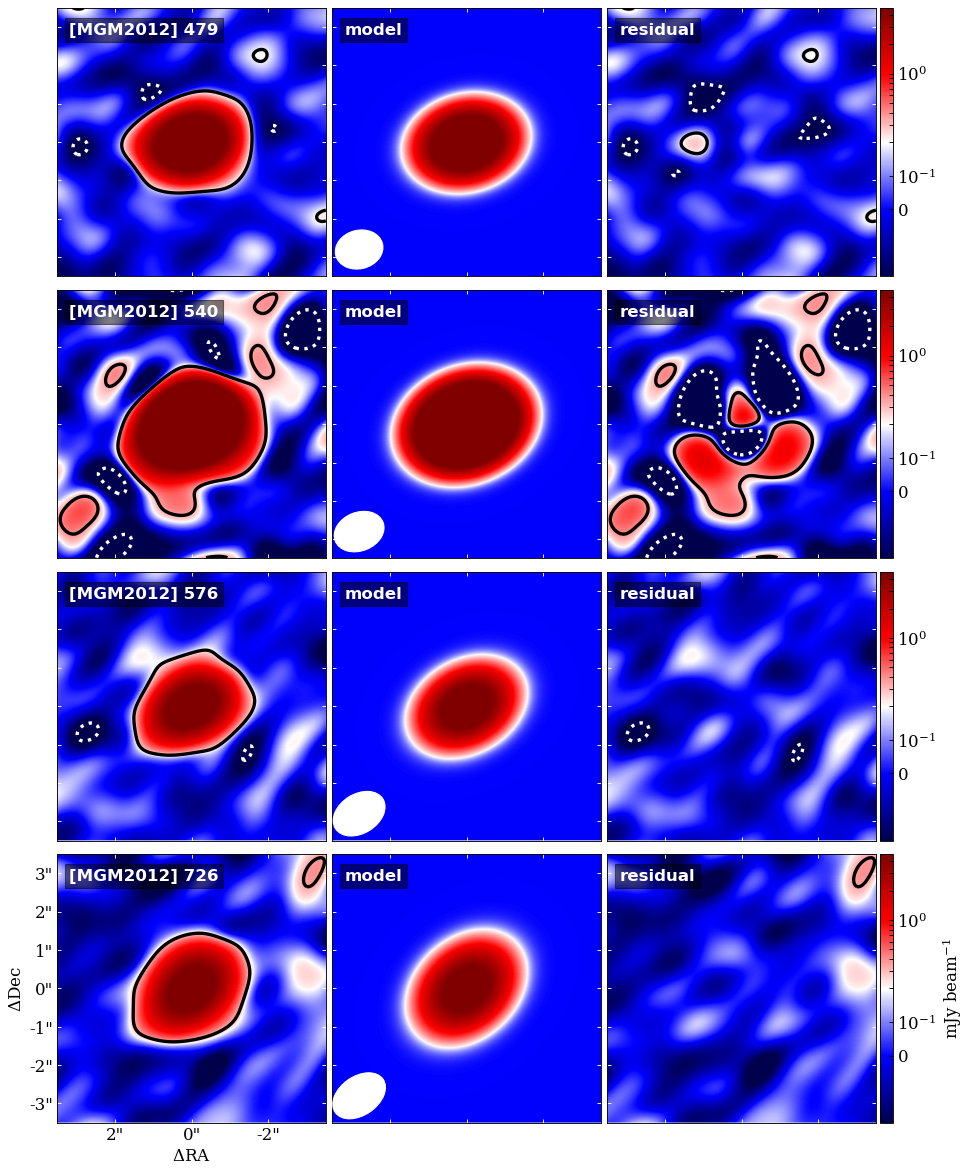}
     \caption{Sources with evidence of resolved emission from double Gaussian fitting, as above.}
     \label{fig:resolved_1}
\end{figure*}

\begin{figure*}
\centering
   \includegraphics[width=16cm]{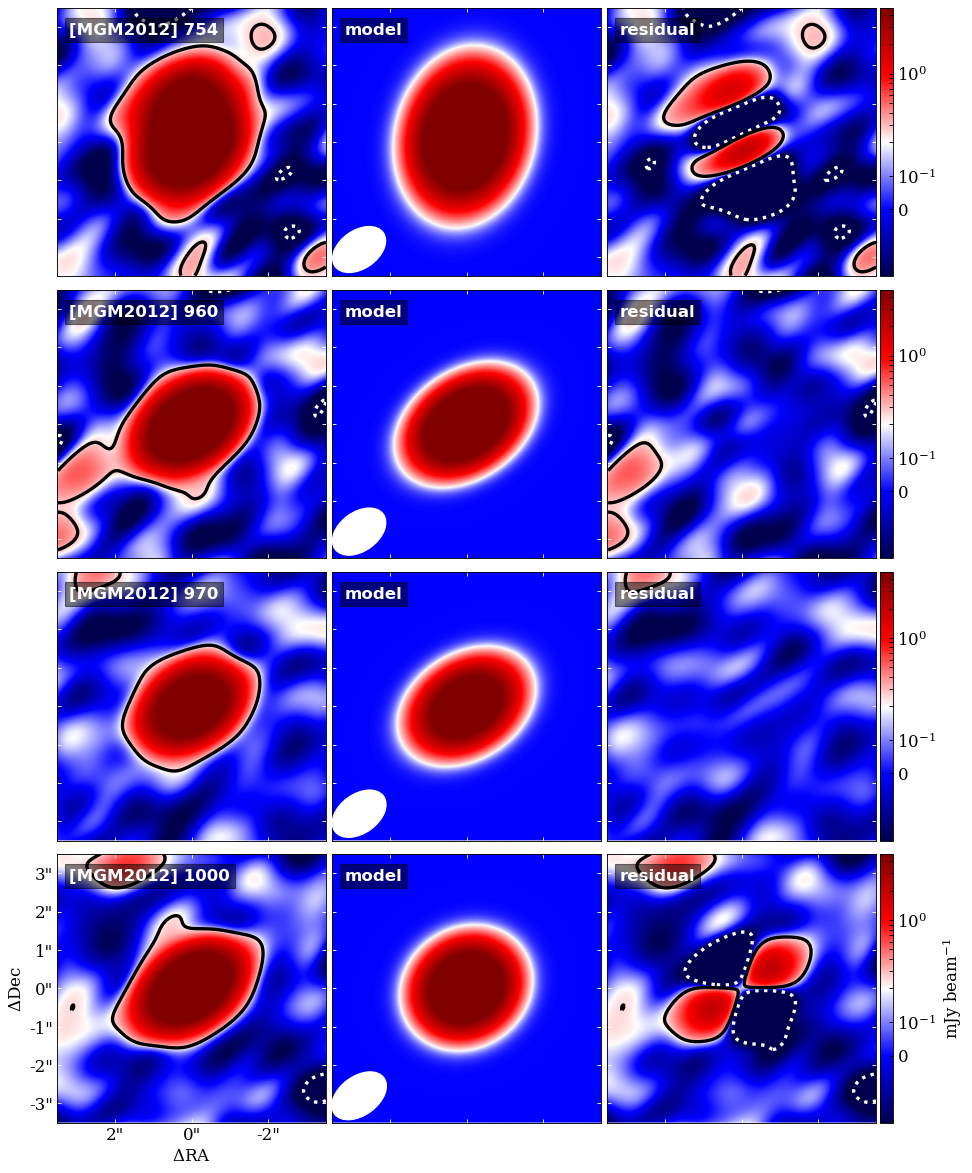}
     \caption{Sources with evidence of resolved emission from double Gaussian fitting, as above.}
     \label{fig:resolved_2}
\end{figure*}

\begin{figure*}
\centering
   \includegraphics[width=16cm]{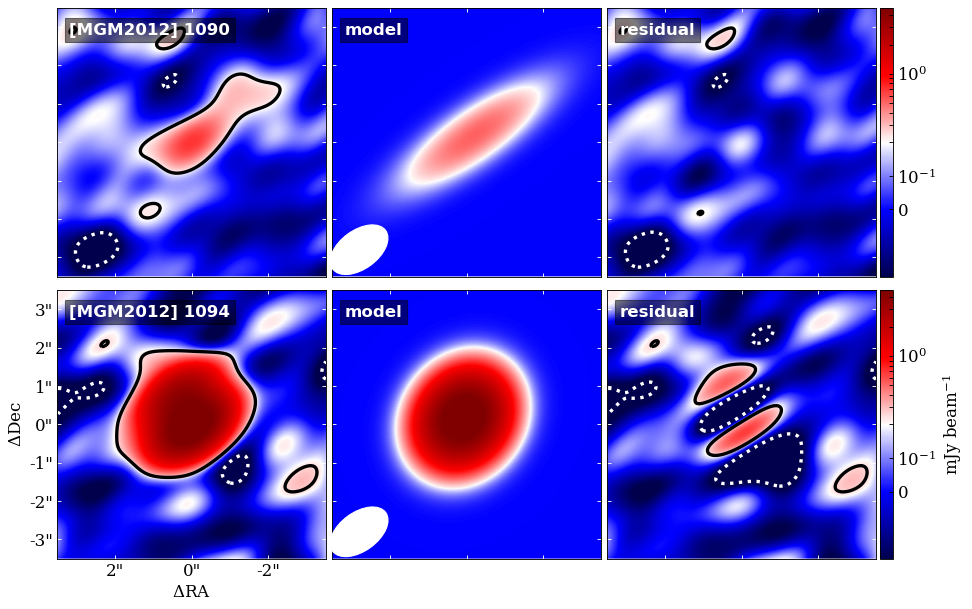}
     \caption{Sources with evidence of resolved emission from double Gaussian fitting, as above.}
     \label{fig:resolved_3}
\end{figure*}

\begin{table*}[b]
	\centering
	\caption{Results from double-Gaussian fits to (marginally) resolved Class II disks in SODA}
	\label{tab:app_fits}
		\begin{tabular}{cccccc}
		\hline \hline \\[-8pt]
		Cluster & $I_{\rm{peak}}$ & $F_{\rm{fit}}$ & $a$ & $b$ & PA \\[2pt]
		 & [mJy]\,[beam$^{-1}$] & [mJy] & [mas] & [mas] & [deg] \\[2pt]
		\hline \\[-8pt]
105 & $5.6 \pm 0.1$ & $7.2 \pm 0.28$ & $29.1 \pm 0.28$ & $20.8 \pm 0.12$ & $142.11 \pm 0.019$\\
129 & $55.6 \pm 0.14$ & $75 \pm 1$ & $32.0 \pm 0.33$ & $21.9 \pm 0.22$ & $98.57 \pm 0.019$\\
371 & $51.3 \pm 0.1$ & $57.6 \pm 0.28$ & $24.7 \pm 0.046$ & $22.3 \pm 0.04$ & $98.07 \pm 0.013$\\
458 & $11.3 \pm 0.06$ & $16.8 \pm 0.19$ & $29.7 \pm 0.13$ & $25.3 \pm 0.11$ & $104.33 \pm 0.019$\\
479 & $17.1 \pm 0.06$ & $18.9 \pm 0.17$ & $27.2 \pm 0.07$ & $20.6 \pm 0.06$ & $110.64 \pm 0.007$\\
540 & $75.1 \pm 0.11$ & $82.3 \pm 0.62$ & $27.7 \pm 0.05$ & $21.5 \pm 0.07$ & $156.08 \pm 0.007$\\
576 & $8.8 \pm 0.08$ & $9.1 \pm 0.10$ & $29.0 \pm 0.14$ & $21.2 \pm 0.11$ & $111.31 \pm 0.011$\\
726 & $7.8 \pm 0.10$ & $9.1 \pm 0.17$ & $31.0 \pm 0.12$ & $23.2 \pm 0.16$ & $20.62 \pm 0.012$\\
754 & $22.9 \pm 0.09$ & $38.8 \pm 0.58$ & $37.1 \pm 0.31$ & $28.3 \pm 0.23$ & $159.82 \pm 0.021$\\
960 & $21.6 \pm 0.09$ & $23.5 \pm 0.14$ & $31.6 \pm 0.05$ & $22.0 \pm 0.08$ & $124.43 \pm 0.005$\\
970 & $13.5 \pm 0.08$ & $15.1 \pm 0.15$ & $31.6 \pm 0.08$ & $22.7 \pm 0.06$ & $148.02 \pm 0.005$\\
1000 & $18.9 \pm 0.08$ & $20.2 \pm 0.16$ & $32.5 \pm 0.05$ & $21.6 \pm 0.08$ & $121.30 \pm 0.004$\\
1090 & $0.7 \pm 0.07$ & $1.2 \pm 0.15$ & $47.0 \pm 4.0$ & $38 \pm 3.1$ & $34.91 \pm 0.31$\\
1094 & $6.5 \pm 0.08$ & $9.6 \pm 0.28$ & $34.8 \pm 0.42$ & $29.6 \pm 0.36$ & $53.40 \pm 0.05$\\
		\hline \\[-8pt]
		\end{tabular}
\end{table*}

\FloatBarrier

\clearpage

\FloatBarrier

\section{The impact of a mixed population of YSOs on the median disk mass}

\begin{figure*}[b]
\centering
   \includegraphics[width=0.95\textwidth]{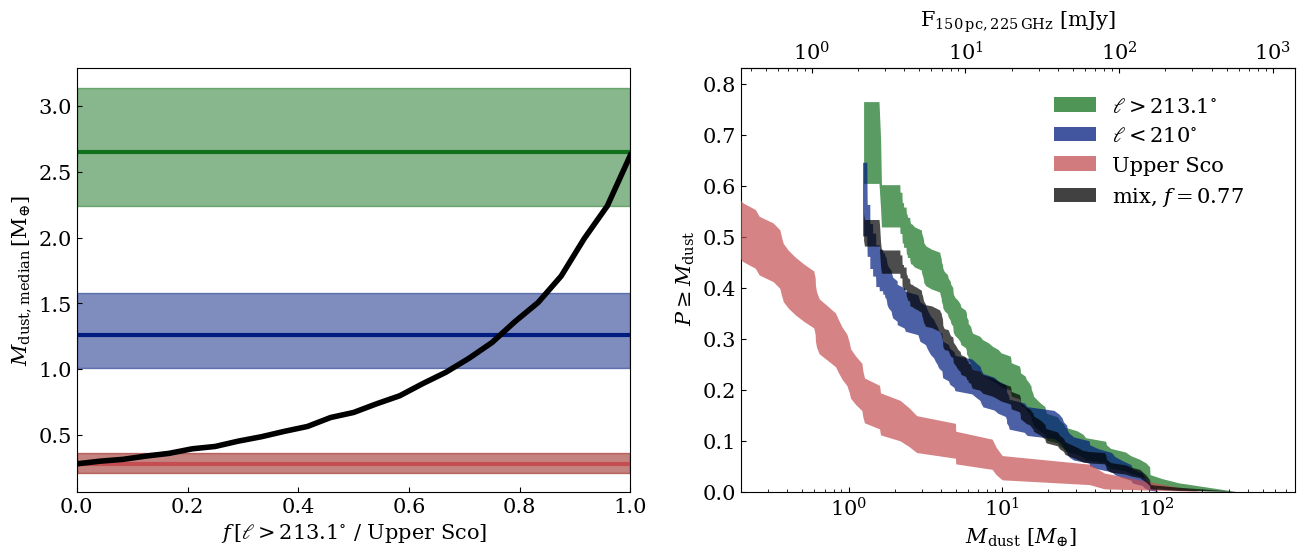}
     \caption{Properties of a mixture of two populations of Class II objects with different ages. {\it Left:} Median disk masses of distributions sampling $N=10^5$ disks with replacement from $\ell > 213.1^{\circ}$ and Upper Sco~\citep{barenfeld16} with ratio $f$ (black), and the median disk masses and uncertainties for Upper Sco (red), SODA between $\ell < 210^{\circ}$ (blue), and between $\ell > 213.1^{\circ}$ (green). {\it Right:} Kaplan-Meier estimators for the disk mass distributions resulting from a draw of 500 samples with $f=0.77$ (dark gray), in SODA between $\ell > 213.1^{\circ}$ (green), $\ell < 210^{\circ}$ (blue), and in Upper Sco (red).}
     \label{fig:popmix}
\end{figure*}

\clearpage
\end{appendix}

\end{document}